\documentclass[5p, number, sort&compress]{elsarticle} 

\usepackage{graphicx}
\usepackage{dcolumn}
\usepackage{bm}
\usepackage{hyperref}
\usepackage{mathtools}
\usepackage{verbatim}
\usepackage{amssymb} 
\usepackage{lineno}

\journal{Nuclear Instruments and Methods}

\begin{document}

\begin{frontmatter}
\title{Track clustering with a quantum annealer for primary vertex reconstruction at hadron colliders}
\author{S. Das\corref{cor1}}
 \ead{souvik@purdue.edu}
\author{A. J. Wildridge}
\author{A. W. Jung}
 \ead{anjung@purdue.edu}
\address{Department of Physics and Astronomy, Purdue University}
\date{\today}
\begin{abstract}
Clustering of charged particle tracks along the beam axis is the first step in reconstructing the positions of hadronic interactions, also known as primary vertices, at hadron collider experiments. We use a 2036 physical qubit D-Wave quantum annealer to perform track clustering in a limited capacity on artificial events where the positions of primary vertices and tracks resemble those measured by the Compact Muon Solenoid experiment at the Large Hadron Collider. The algorithm, which is not a classical-quantum hybrid but relies entirely on quantum annealing, is tested on a variety of event topologies. We demonstrate a deterministic graph-embedding of the problem on the D-Wave Chimera architecture, a method for optimizing the coupling strengths within logical qubits, and a method for optimizing annealing time. Further, we benchmark it against simulated annealing on a commercial CPU constrained to the same processor time per anneal as the physical annealer. We note a quantum advantage against simulated annealing up to a 56 logical qubit problem that involves 665 physical qubits on average. Our embedding and optimization methods, and the benchmarking paradigm, can be applied generally to other clustering problems on quantum annealers. This algorithm may be used as a building-block for more sophisticated algorithms to reach the number of primary vertices at the LHC.
\end{abstract}

\end{frontmatter}

\section{\label{sec:Introduction}Introduction}

\begin{figure*}
\centering
\includegraphics[width=0.8\linewidth]{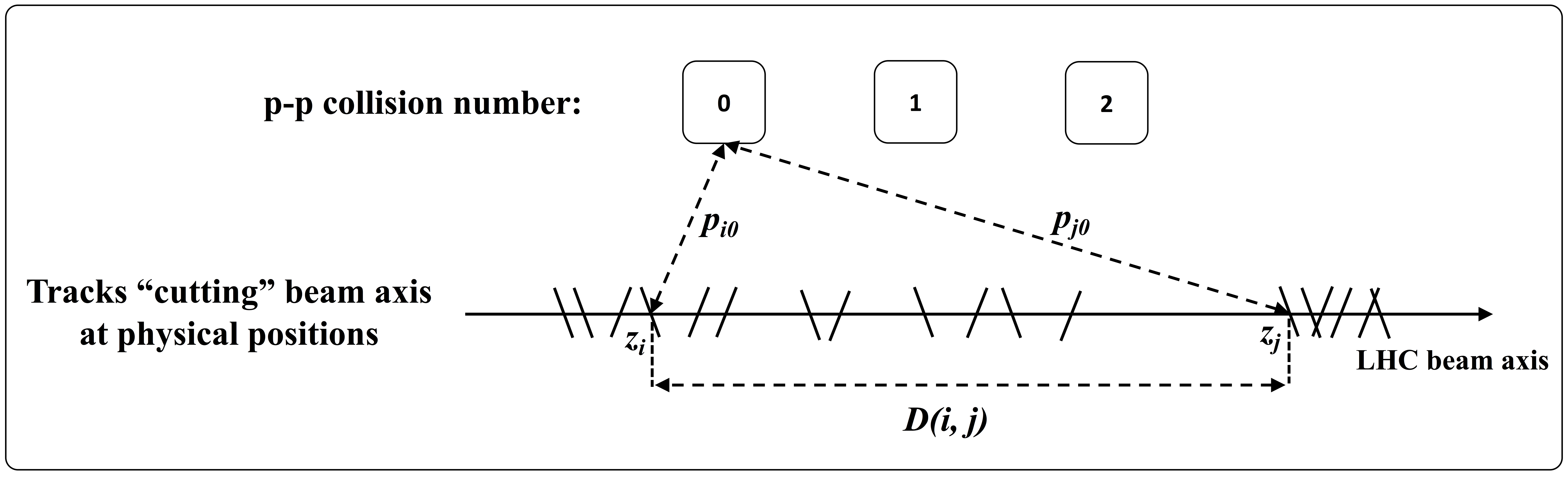} 
\caption{Illustration of the penalty $D(i,j)$ imposed by the problem Hamiltonian in Eq.~\ref{eq:ObjectiveFunction} when widely separated tracks in $z_0$, $z_i$ and $z_j$ are associated with the same p-p collision. Collisions are labeled by integers from 0 to $n_V - 1$. The algorithm solves for the association matrix $p_{ik}$ where $i$ is the track index and $k$ is the collision index.}
\label{fig:Illustration}
\end{figure*}

\subsection{Primary Vertex Reconstruction}

Hadron colliders circulate counter-rotating beams of hadrons in closely packed bunches that cross at designated interaction points. These interaction points are instrumented with experiments that detect particles produced at hadron-hadron collisions when the bunches cross. Reconstructing the positions of these collisions within a bunch crossing, also known as primary vertices, from the trajectories of charged particles detected by the apparatuses is a crucial step in the process of fully reconstructing a collision event based on recorded detector information. Upon completion, the event reconstruction yields observables such as particle momenta, energies and other quantities~\cite{CMSParticleFlow} which are of paramount importance for data analyses. Primary vertex reconstruction accuracy affects the precise calculation of particle directions in three dimensions, rejection of hadronic jets that do not originate at the collision of interest with high momentum exchange between partons, and the determination of displaced vertices of jets that originate from bottom quarks. The Large Hadron Collider (LHC) is a high luminosity collider that produced an average of 35 proton-proton (p-p) collisions at each bunch crossing in 2018~\cite{CMSPileup}. Typical numbers for the high-luminosity phase of the LHC (HL-LHC) are expected to be approximately 200~\cite{HLLHCTDR}. These collisions are distributed widely in one dimension along the beam axis with a minute but non-zero spread in the transverse direction. The transverse spread is ignored in this study. At one of the LHC interaction points, the Compact Muon Solenoid experiment (CMS) reconstructs the paths of charged particles from p-p collisions as tracks detected by its silicon tracker~\cite{Chatrchyan:2008aa}. Track reconstruction algorithms propagate detector uncertainties which obscure which tracks originated together at a primary vertex. Thus, primary vertex reconstruction begins with a one-dimensional clustering of tracks by their positions along the beam axis where they approach it most closely, also known as the tracks' $z_0$. This clustering is done on a classical computer using annealing algorithms that mimic the approach of a physical system to its lowest energy state iteratively through a series of cooling operations at both the CMS and ATLAS experiments at the LHC~\cite{CMSTracking, ATLASPrimaryVertexing}. In this paper, we demonstrate a method of performing this clustering in one physical step on a D-Wave quantum annealer and report preliminary results benchmarked against simulated annealing on a classical computer. While primary vertex reconstruction is not among the most computationally intensive tasks, this work is an early example of a high energy physics application that has been implemented not on a simulated quantum system, or using a classical-quantum hybrid algorithm, but purely as a quantum annealing algorithm on a physical device. As such, this work is a proof-of-principle limited by current technology that represents a step forward along the path of realizing practical quantum computing for high energy physics in particular and science in general. This algorithm, as is, may be used as a building-block for more sophisticated algorithms to reach the number of primary vertices relevant for the LHC. See Refs.~\cite{Mott:2017Nature, Zlokapa:2019tkn, Cormier:2019kcq, Wei:2019rqy, Bapst:2019} for similar exploratory applications of quantum annealing in high energy physics.

\subsection{\label{sec:Annealer}The D-Wave Quantum Annealer}

The D-Wave 2000Q\_6 quantum processor, available from D-Wave Systems Inc., performs computations through quantum annealing \cite{Finnila:1994, Kadowaki:1998, Santoro:2002}. The quantum processing unit (QPU) has 2036 RF-SQUID flux qubits implemented as superconducting niobium loops \cite{Harris:2010}. Each qubit has a programmable external magnetic field to bias it. The network of qubits is not fully connected and programmable couplings have been implemented between 5967 pairs of qubits. A computational problem is defined by setting the biases ($h_i$) and couplings ($J_{ij}$) such that the ground state of the qubits' Hamiltonian corresponds to the solution. We call this the ``problem Hamiltonian" ($H_p$)
\begin{equation}
    H_p = \sum_{i} h_i \sigma_z^i + \sum_{i} \sum_{j > i} J_{ij} \sigma_z^i \sigma_z^j,
\label{eq:ProblemHamiltonian}
\end{equation}
where $\sigma_z^i$ is a spin projection observable of the $i^\mathrm{th}$ qubit with eigenvalues $+1$ and $-1$. (This $z$ direction is not related to the beam axis at CMS.) It may be trivially mapped to a bit observable $q_i$ with eigenvalues 0 and 1 through the shift $2 q_i = \sigma_z^i + I$, where $I$ is the identity matrix. The problem Hamiltonian may then be expressed for quadratic unconstrained binary optimization (QUBO) in the dimensionless form
\begin{equation}
    Q_p = \sum_{i} a_i q_i + \sum_{i} \sum_{j > i} b_{ij} q_i q_j + \mathrm{constant},
\label{eq:ProblemHamiltonianQUBO}
\end{equation}

At the beginning of a typical annealing cycle in the QPU, a driver Hamiltonian puts all qubits in a superposition of the computational basis states by introducing a global energy bias in the transverse $x-$direction. Annealing proceeds by lowering this driver Hamiltonian while simultaneously increasing the problem Hamiltonian as
\begin{equation}
    H = A(s) \sum_i \sigma_x^i + B(s) Q_p,
\label{eq:Annealing}
\end{equation}
where $A$ is a monotonically decreasing function and $B$ is a monotonically increasing function defined on $s \in [0, 1]$. $A$ and $B$ have units of energy. The adiabatic theorem of quantum mechanics guarantees that the qubits will land in the minimum of $Q_p$ (which is the ground state of $H_p$) if this change is sufficiently gradual, the ground state is unique with a non-zero energy separation from other states, and the initial state of the qubits is the ground state of the initial field \cite{BornFock:1928, Kato:1950, AvronElgart:1999}. These conditions are difficult to achieve experimentally. We therefore anneal within tens of microseconds during which quantum tunneling leaves the system in a low energy configuration at the end of the annealing process~\cite{ThermallyAssistedQuantumAnnealing}. We measure the final state of the qubits as a solution, and repeat several times. The lowest energy solution is then taken as the best one.

In programming the QPU, a set of physical qubits are strongly coupled to create a logical qubit in order to compensate for the limited connectivity of a single physical qubit and to mitigate bit flips from thermal fluctuations. The computational problem is expressed over logical qubits and this has to be mapped using graph embedding to the network of physical qubits.

\subsection{Main Contributions}

\begin{enumerate}

\item
We map track clustering to a QUBO form that can be solved by the D-Wave quantum annealer. A generic version of this algorithm has been described in Ref.~\cite{BAHPaper}. We distort the form to encourage annealing to the ground state given the clustering characteristics of tracks observed at CMS.

\item
We use a deterministic graph embedding algorithm from logical qubits to the graph of physical qubits for the QUBO form for clustering. This results in improvements over default D-Wave stochastic embedding algorithms~\cite{CaiMacready, VickyChoi1, VickyChoi2}.

\item
We optimize the chain strengths between the physical qubits that make up a logical qubit. A novel result is obtained that optimal chain strength grows linearly with chain length.

\item
We optimize the annealing time to obtain the shortest time to reach a solution.

\item
We benchmark it against simulated annealing and deterministic annealing running on a commercial CPU. We propose a general benchmarking method based on allowing both the QPU and the CPU to run for the same period of time.

\item
We show how our method scales with event complexity by testing it on a variety of event topologies, where the positions of vertices and tracks are drawn from measured distributions at CMS.

\end{enumerate}

The algorithm may be used hierarchically, to first cluster tracks into two vertices and then sub-cluster them into more vertices, to solve for the number of primary vertices expected at the LHC (35 to 200). However, such extensions bring additional layers of complexity and are beyond the scope of this proof-of-principle paper that focuses on the algorithmic primitive.

\section{\label{sec:Formulation}Formulation}

For track clustering, we seek the minimum of the problem QUBO form
\begin{equation}
\begin{split}
    Q_p = \sum_k^{n_V} \sum_i^{n_T} \sum_{j > i}^{n_T} p_{ik}p_{jk} g(D(i, j); m) \\
    + \lambda \sum_i^{n_T} \left( 1 - \sum_k^{n_V} p_{ik} \right)^2,
\label{eq:ObjectiveFunction}
\end{split}
\end{equation}
where $n_T$ is the number of tracks, $n_V$ is the number of vertices, $p_{ik} \in \left[0, 1\right]$ is the probability of the $i^\mathrm{th}$ track to be associated with the $k^\mathrm{th}$ vertex, $g$ is a distortion function, $m$ is a distortion parameter, $D(i, j)$ is a measure of distance between the reconstructed $z_0$ parameters of the $i^\mathrm{th}$ and $j^\mathrm{th}$ tracks, and $\lambda$ is a penalty parameter.

As in any clustering algorithm, a density threshold of tracks must be set that determines $n_V$. For the study presented in this paper, we order the tracks in $z_0$ and count the number of gaps greater than a threshold of 5 mm. The number of gaps plus one is equal to the number of vertices, $n_V$. In general, this algorithm could be applied on an event with unknown numbers of vertices by scanning the number of vertices around that implied by the nominal track density and considering the optimal solution from the Elbow method~\cite{elbow}, the Akaike Information Criterion, the Bayesian Information Criterion~\cite{AICBIC}, or the Silhouette Method~\cite{silhouette}.

\begin{figure}
\centering
\includegraphics[width=\linewidth]{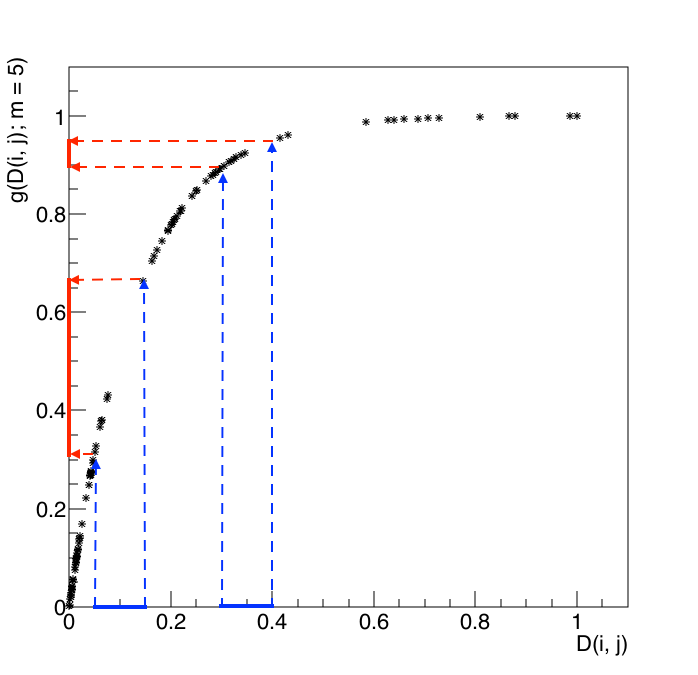}
\caption{The distortion function $g(x; m = 5)$, defined in Eq.~\ref{eq:ExponentialSqueezing}, applied to $D(i, j)$ defined in Eq.~\ref{eq:Dij}. This raises and increases spacing between small values of $D(i,j)$ to mitigate their sensitivity to coupler noise.}
\label{fig:c_exponentialSqueeze}
\end{figure}

For $D(i, j)$, we find the absolute distance between $z_i$ and $z_j$ divided by the quadrature sum of the measurement uncertainties $\delta z_i$ and $\delta z_j$ to be an effective measure:
\begin{equation}
    D(i, j) = \frac{|z_i - z_j|}{\sqrt{\delta z_i^2 + \delta z_j^2}}.
\label{eq:Dij}
\end{equation}
With an increasing numbers of tracks around vertices, $D(i, j)$ tends to cluster near zero. In order to not drown this term by noise in the coupling between qubits, we distort it using
\begin{equation}
    g(x; m) = 1 - e^{-mx},
\label{eq:ExponentialSqueezing}
\end{equation}
where $m$ is a distortion parameter. This raises and increases spacing between small values of $D(i, j)$, which are most susceptible to noise, without changing their order as shown in Fig.~\ref{fig:c_exponentialSqueeze}. $m$ is set to 5 for event topologies considered in this study.

A penalty parameter, $\lambda$, is introduced to discourage $p_{ik}$ for each track to add up to anything other than 1. While it should be large enough to discourage the probability of a single track to be assigned to multiple vertices, it must not drown out the energy scale of $D(i, j)$. We tried several values of $\lambda$ from 1.0 to 2.0 times the maximum of $D(i, j)$ and settled on 1.2 times the maximum of $D(i, j)$ for optimal performance. Not all solutions from the QPU have $p_{ik}$ add up to 1 for all tracks; these are checked offline by a CPU and marked as invalid.

If each $p_{ik}$ is represented by one logical qubit in the QPU, Eq.~\ref{eq:ObjectiveFunction} is already in the QUBO form of the problem Hamiltonian described in Eq.~\ref{eq:ProblemHamiltonianQUBO}. Therefore, it can be directly programmed onto a D-Wave QPU. We need to program $n_V n_T$ logical qubits and $n_V n_T (n_V + n_T - 2) /2 $ couplings between them to encode $H_p$. 

\section{\label{sec:TrackMC}Track Monte Carlo Generation}

\begin{figure}
\centering
\includegraphics[width=\linewidth]{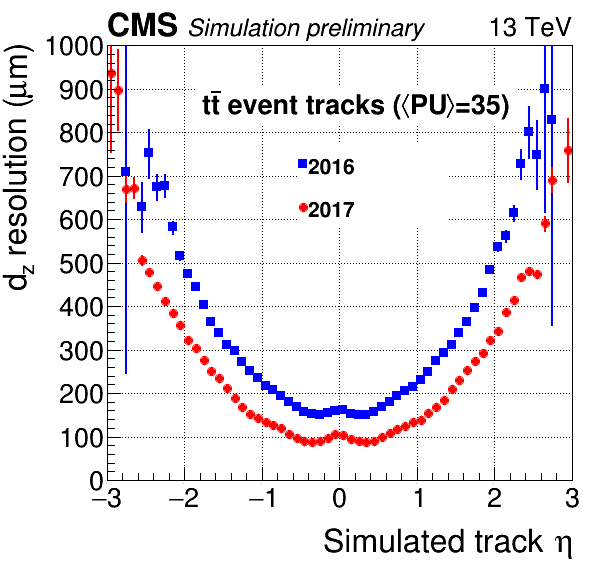}
\caption{Resolution of longitudinal impact point of tracks as a function of the simulated track $\eta$ for CMS 2016 and 2017 detectors. Obtained from publicly available document, \url{https://twiki.cern.ch/twiki/bin/view/CMSPublic/TrackingPOGPerformance2017MC}. The 2017 distribution is used to throw simplified artificial events for this study.}
\label{fig:dzResolution}
\end{figure}

To test and optimize the algorithm, we generate simplified artificial events that each contain a set of tracks coming from primary vertices. The primary vertices lie in one dimension sampled from a Gaussian of width 35 mm to emulate the distribution seen in CMS. The $z_0$ parameter of toy tracks are sampled from Gaussians centered around the generated vertices with widths corresponding to track resolutions published in CMS~\cite{CMSTracking}. These widths range from 100 to 700 $\mu$m depending on the pseudorapidity ($\eta$) of the tracks, as shown in Fig.~\ref{fig:dzResolution}. Our set of simplified artificial events containing primary vertex positions and $z_0$ of tracks are published in~\cite{das_souvik_2020_3786899}.

Our QUBO formulation cannot directly embed the approximately 35 primary vertices typically produced in p-p collisions at the LHC onto the D-Wave 2000Q\_6 processor. In this study, we choose to study seven simple event topologies. In ascending order of the number of logical qubits needed to formulate them, they are: 2 vertices and 10 tracks, 2 vertices 16 tracks, 3 vertices 15 tracks, 2 vertices 28 tracks, 4 vertices 16 tracks, 5 vertices 15 tracks, and 4 vertices 20 tracks. We tabulate the number of logical qubits and couplers, and the average chain length determined by the embedding algorithm, for each topology in Table~\ref{tab:Table_SA}. As expressed in our Conclusions~\ref{sec:Conclusions}, our formulation may be used recursively to solve event complexities relevant for the LHC.

\section{\label{sec:Embedding}Deterministic Embedding of the Formulation}

\begin{figure}
\centering
\includegraphics[width=\linewidth]{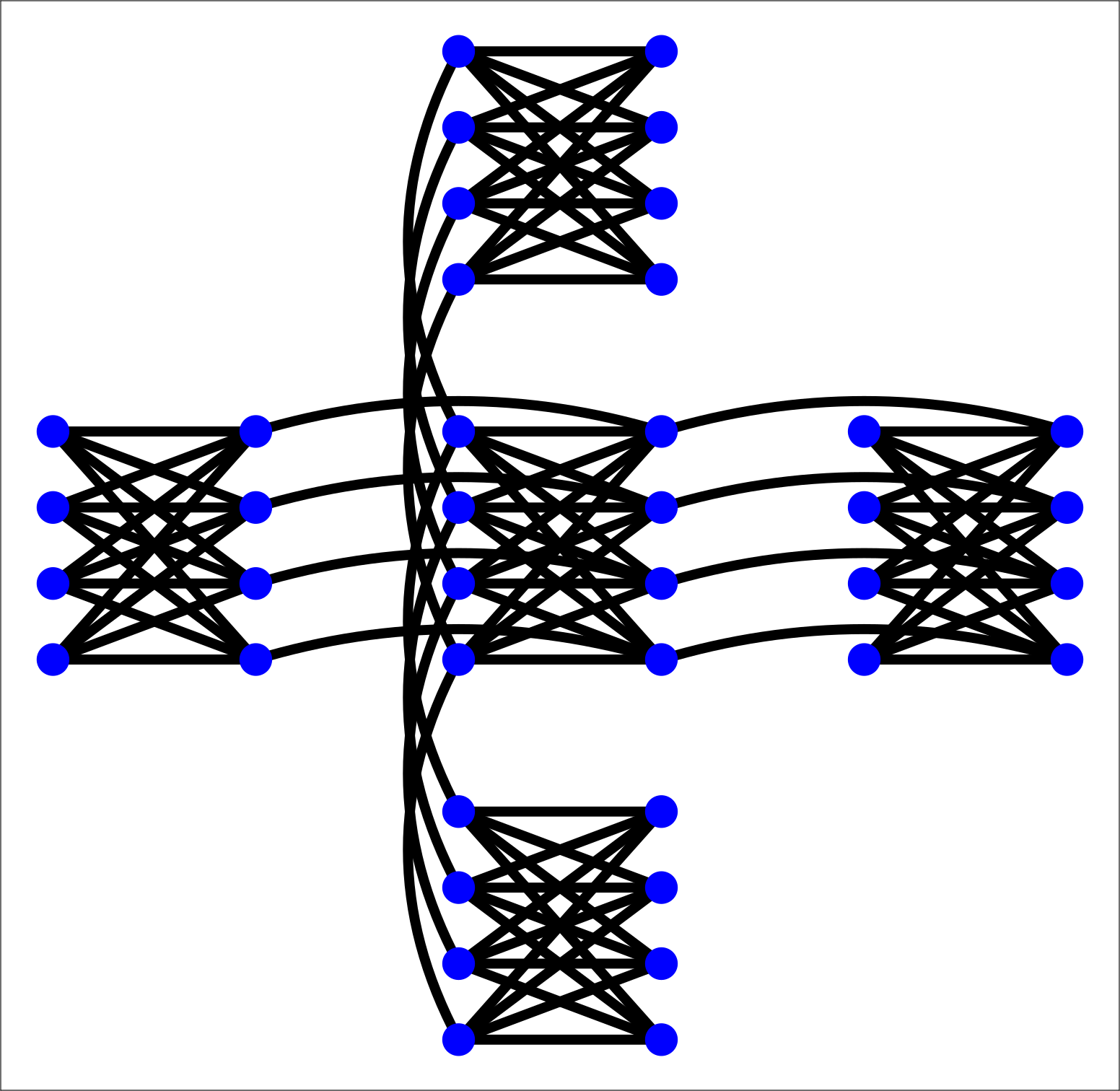}
\caption{A unit cell of 8 physical qubits in the Chimera graph surrounded by 4 other unit cells illustrates the general connectivity pattern in the D-Wave 2000Q\_6 architecture. Each physical qubit of the central unit cell has 6 couplers.}
\label{fig:Chimera}
\end{figure}

\begin{figure*}
\centering
\includegraphics[width=0.8\linewidth]{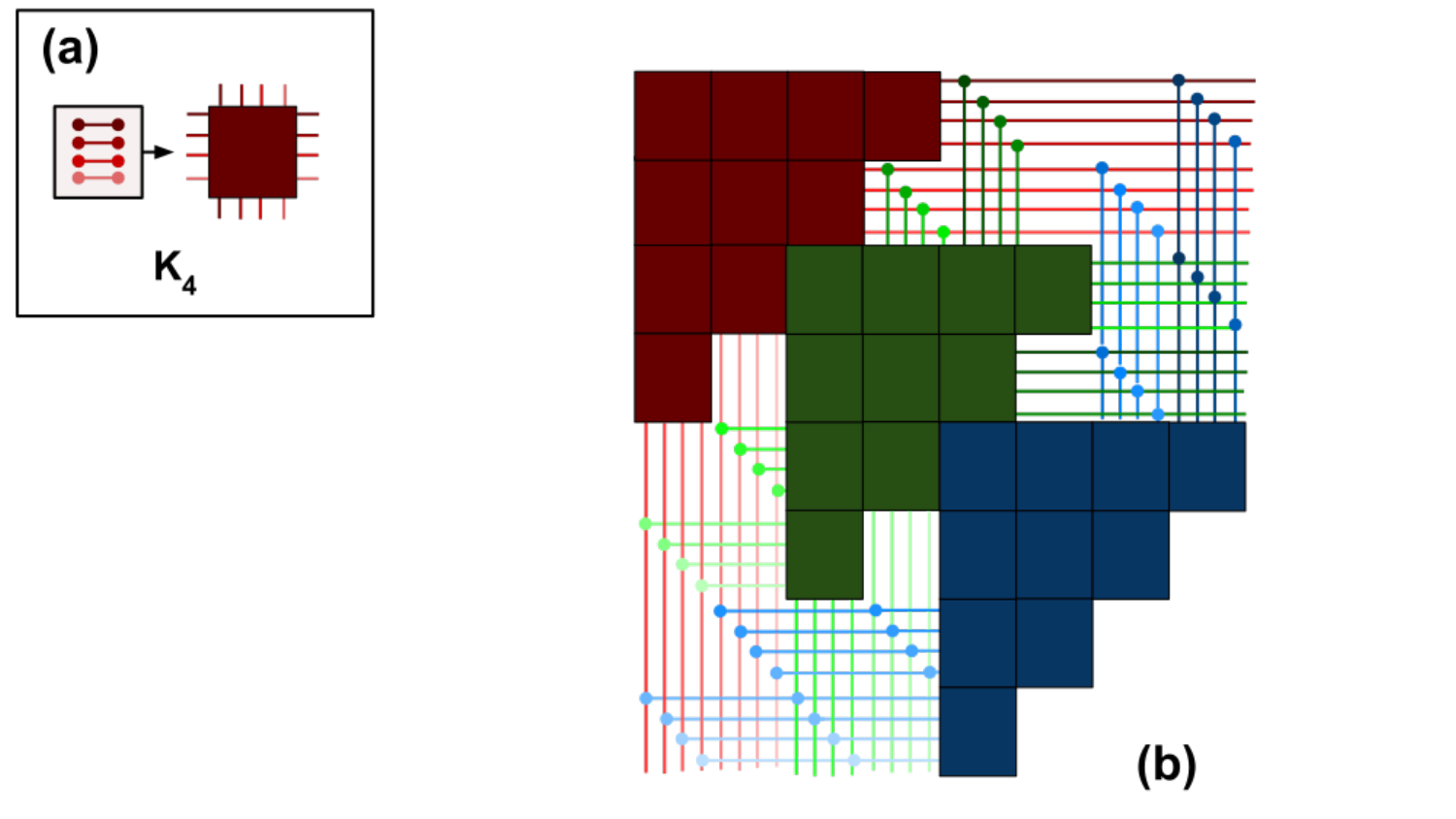} 
\caption{Illustration of the Cartesian product of fully connected graphs embedded on a Chimera graph. a) Each rectangular box corresponds to a unit cell with 8 qubits. b) $K_3 \Box K_{15}$, corresponding to 3 vertices 15 tracks.}
\label{fig:NexusBusConnections}
\end{figure*}

\begin{figure}
\centering
\includegraphics[width=\linewidth]{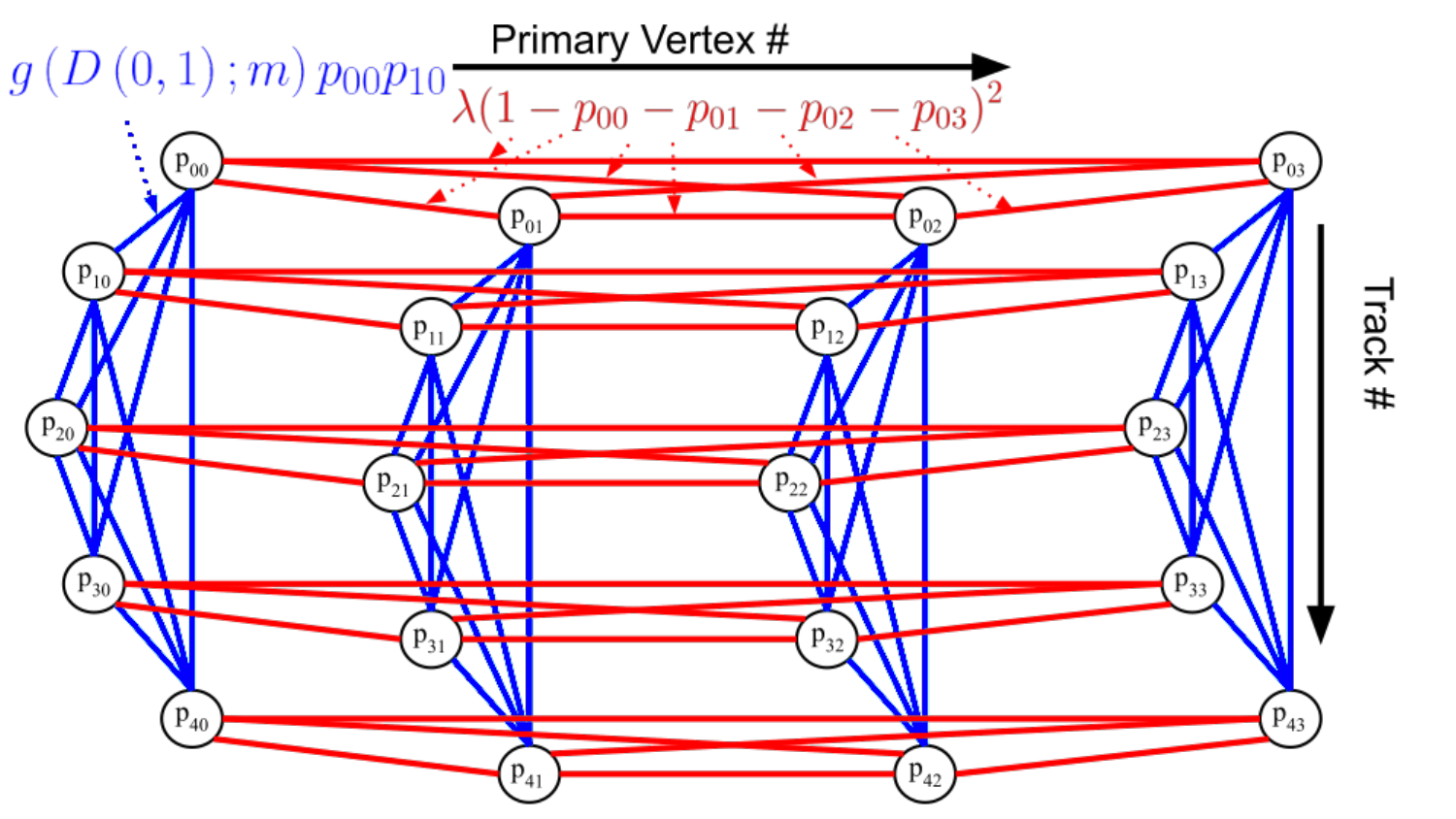}
\caption{Illustration of the fully-connected Cartesian product nature of the problem Hamiltonian expressed in Eq.~\ref{eq:ObjectiveFunction}}
\label{fig:CartesianProduct}
\end{figure}

\begin{figure}
\centering
\includegraphics[width=\linewidth]{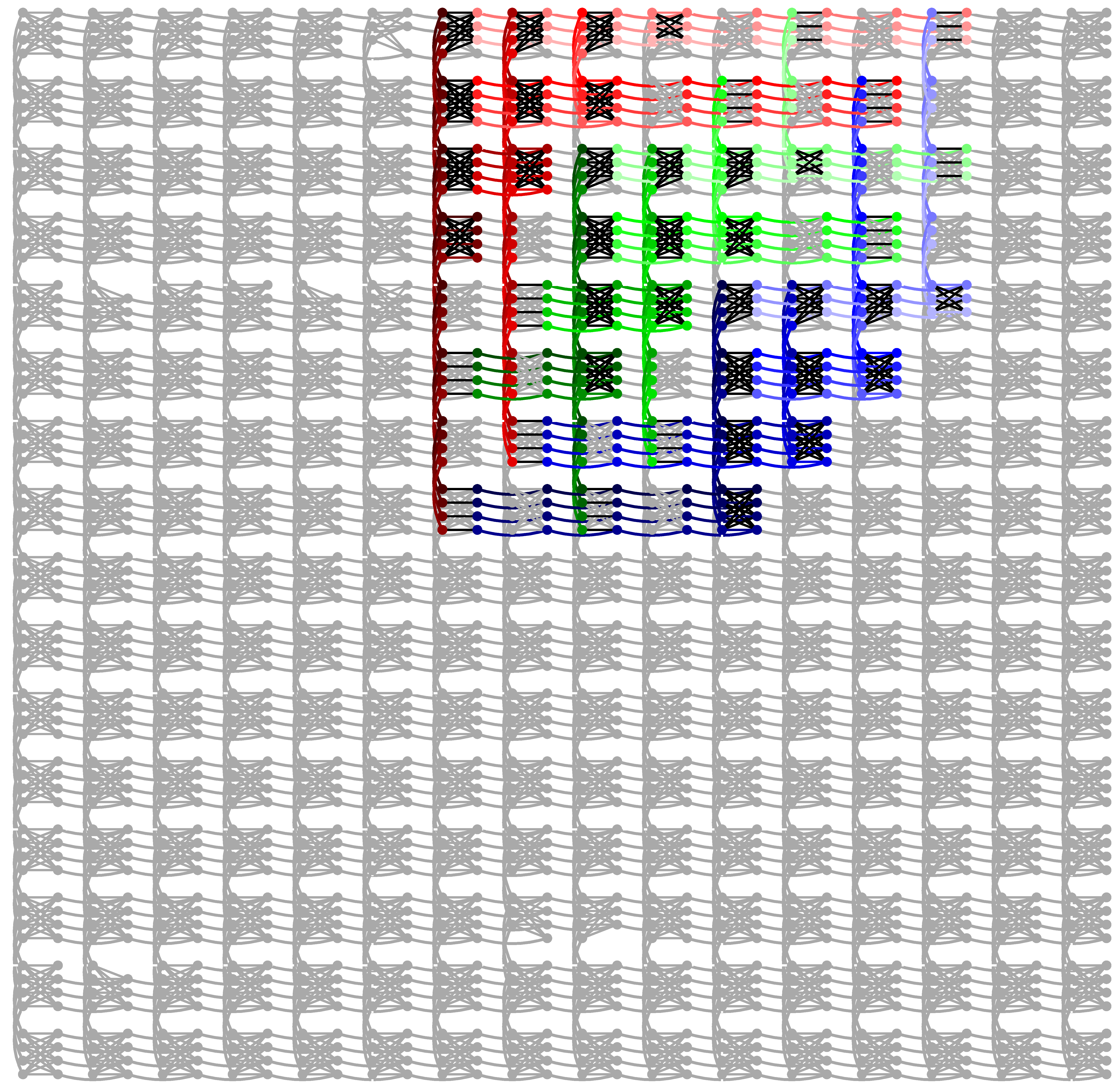}
\caption{Representation of the deterministic embedding described in Section~\ref{sec:Embedding} on the 2000Q\_6 QPU for the problem-size of 3 vertices 15 tracks.}    
\label{fig:3V_15T_embedding}
\end{figure}

The connectivity graph of physical qubits in the D-Wave 2000Q\_6 QPU is known as the Chimera architecture. A unit cell of the Chimera architecture consists of 8 physical qubits coupled in a bipartite graph. In Fig.~\ref{fig:Chimera}, we depict such a unit cell surrounded by 4 other unit cells to illustrate the connectivity pattern. Each physical qubit of the central unit cell has 6 couplers. Unit cells are tiled in a 16 $\times$ 16 matrix in the 2000Q\_6 QPU.

Our formulation of track clustering as described in Eq.~\ref{eq:ObjectiveFunction} finds a natural embedding in the Chimera architecture because it can be seen as a Cartesian product of two fully-connected graphs as we illustrate in Fig.~\ref{fig:CartesianProduct}. Hence, we use the results of Ref.~\cite{DeterministicEmbedding} to embed our formulation efficiently and deterministically in the Chimera architecture. In Fig.~\ref{fig:NexusBusConnections} we illustrate this embedding schematically with 3 vertices 15 tracks. Each rectangular box corresponds to a unit cell with 8 qubits. The physical qubits within a box are not connected across rows. The result of this embedding on the 2006\_6 QPU is shown in Fig.~\ref{fig:3V_15T_embedding}.

Real quantum annealers like the 2000Q\_6 do not possess perfect yields on qubits and couplers. The deterministic embedding algorithm is flexible enough to handle missing qubits and couplers.

\begin{table*}
\centering
\begin{tabular}{c c c c c c c}
\hline
\hline
                     &                  & Average & Optimal  & Optimal       &\multicolumn{2}{c}{\underline{Simulated Annealing}} \\
Topology             & Logical          & Chain   & Chain    & Annealing     & time/sweep & Allowed \\
                     & Qubits, Couplers & Length  & Strength & time ($\mu$s) & ($\mu$s)   & Sweeps  \\
\hline
2 vertices 10 tracks & 20, 100          & 5.80    & 1.350    & 45            & 0.6        &  75 \\
2 vertices 16 tracks & 32, 256          & 6.75    & 1.400    & 45            & 1.2        &  38 \\
3 vertices 15 tracks & 45, 360          & 8.64    & 1.500    & 45            & 1.7        &  27 \\
2 vertices 28 tracks & 56, 784          & 11.88   & 1.675    & 80            & 2.7        &  30 \\
4 vertices 16 tracks & 64, 576          & 10.63   & 1.600    & 80            & 2.5        &  32 \\
5 vertices 15 tracks & 75, 675          & 13.15   & 1.750    & 80            & 3.1        &  26 \\
4 vertices 20 tracks & 80, 880          & 15.56   & 1.875    & 80            & 3.5        &  23 \\
\hline
\hline
\end{tabular}
\caption{For each of the studied topologies, we tabulate the number of logical qubits and couplers described in Section~\ref{sec:Formulation}. The average chain length is determined by the embedding algorithm in Section~\ref{sec:Embedding}. The optimal chain strength is determined in Section~\ref{sec:ChainStrengthOptimization}. The optimal QPU annealing time is determined in Section~\ref{sec:AnnealingTimeOptimization}. The time taken per sweep by simulated annealing on a commercial CPU is measured in Section~\ref{sec:Benchmark_SA}. The number of allowed sweeps is the annealing time divided by the time/sweep and rounded to a whole number.}
\label{tab:Table_SA}
\end{table*}

\begin{table*}
\centering
\begin{tabular}{c c c c}
\hline
\hline
                     & Optimal       &\multicolumn{2}{c}{\underline{Deterministic Annealing}} \\
Topology             & Annealing     & time/sweep & Allowed  \\
                     & time ($\mu$s) & ($\mu$s)   & Sweeps \\
\hline
2 vertices 10 tracks & 45            & 0.9          & 50 \\
2 vertices 16 tracks & 45            & 1.1          & 41 \\
3 vertices 15 tracks & 45            & 1.8          & 25 \\
2 vertices 28 tracks & 80            & 1.8          & 44 \\
4 vertices 16 tracks & 80            & 3.3          & 24 \\
5 vertices 15 tracks & 80            & 2.8          & 29 \\
4 vertices 20 tracks & 80            & 2.8          & 29 \\
\hline
\hline
\end{tabular}
\caption{For each studied topology, we measure the time taken per sweep by deterministic annealing on a commercial CPU as described in Section~\ref{sec:Benchmark_DA}. The number of allowed sweeps is the annealing time divided by the time/sweep and rounded to a whole number.}
\label{tab:Table_DA}
\end{table*}

\section{\label{sec:ChainStrengthOptimization}Chain Strength Optimization}

\begin{figure}
\centering
\includegraphics[width=\linewidth]{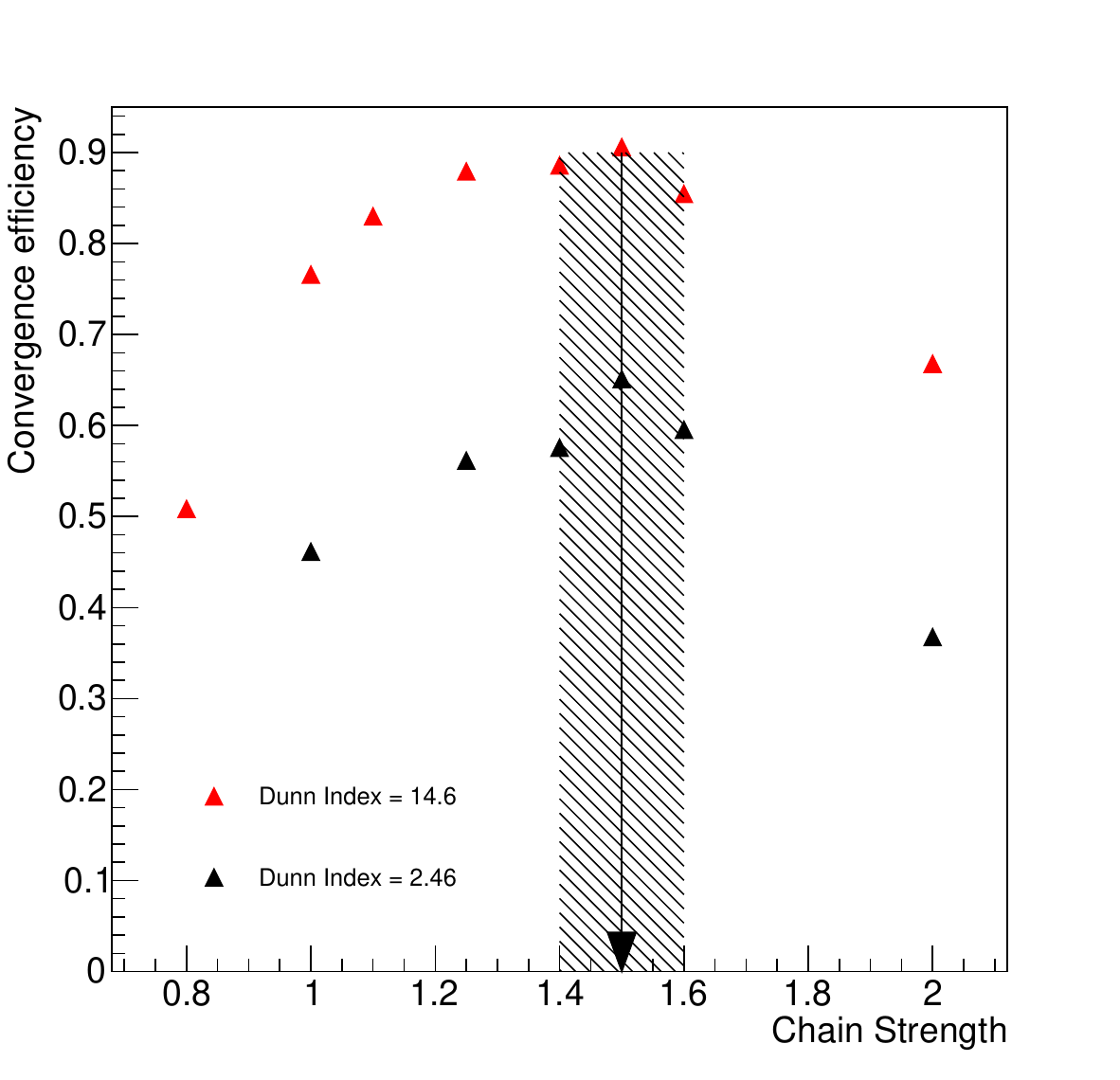}
\caption{Chain strength scan for two events containing 3 primary vertices and 15 tracks with different Dunn Indices. Statistical uncertainties are shown and are negligible due to large sample sizes. The optimal chain strength is found to be $1.5 \pm 0.1\ h$GHz.}
\label{fig:3V_15T_chain_strength}
\end{figure}

\begin{figure}
\centering
\includegraphics[width=\linewidth]{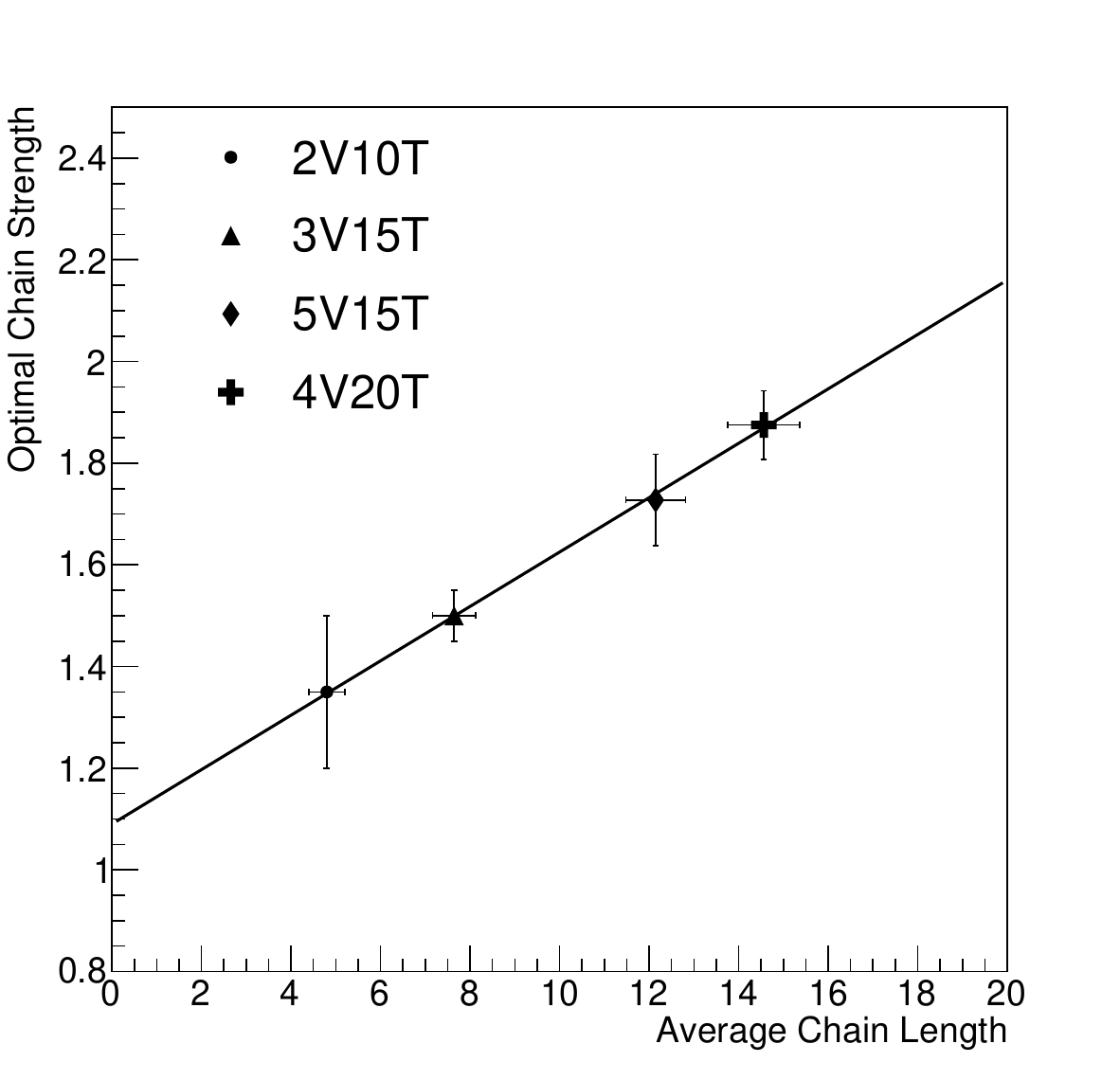}
\caption{Optimal chain strength for 4 event topologies as a function of average chain length within a logical qubit. Uncertainty in average chain length is the statistical standard error on the mean. Uncertainty in the optimal chain strength is described in Section~\ref{sec:ChainStrengthOptimization}. A straight line fits the points and is used to predict the optimal change length for arbitrary topologies.}
\label{fig:optimal_cs}
\end{figure}

Having deterministically embedded track clustering to the Chimera architecture, we optimize the coupling strength between the physical qubits of a logical qubit, also known as chain strength. High chain strengths encourage the physical qubits to align, but when too high can dominate the problem Hamiltonian and cause early freeze-out of the qubits to sub-optimal solutions. Too low chain strengths, on the other hand, can be overwhelmed by the problem Hamiltonian and result in broken chains and random bit flips. Therefore, there exists an optimal chain strength. 

We scan chain strength and observe its effect on the convergence efficiency as shown in Fig.~\ref{fig:3V_15T_chain_strength}. We find that the optimal strength is independent of the ``clumpiness" of the data represented by its Dunn index (Eq.~\ref{eq:Dunn}). Convergence efficiency, defined as the ratio of the number of correct solutions to the number of times the QPU is sampled, is determined by running our formulation with the D-Wave QPU on artificial data that is generated as discussed in Section~\ref{sec:Results}. Each point represents 10,000 samples of the QPU and hence the statistical uncertainties are small. Optimal chain strength occurs at the highest convergence efficiency. The uncertainty in the optimal chain strength is determined by finding the furthest neighboring points that are lower by one standard deviation. For 3 vertices and 15 tracks, the optimal chain strength is found to be $1.5 \pm 0.1\ $ DAC counts.

By plotting the optimal chain strength for multiple event topologies in Fig.~\ref{fig:optimal_cs}, we obtain the result that it is linear with respect to the average chain length. More complex event topologies involving more vertices and tracks result in longer chains of physical qubits for a logical qubit. Hereafter, we use this linearity to predict the optimal chain strength of any arbitrary topology.

\section{\label{sec:AnnealingTimeOptimization}Annealing Time Optimization}

\begin{figure}
\centering
\includegraphics[width=\linewidth]{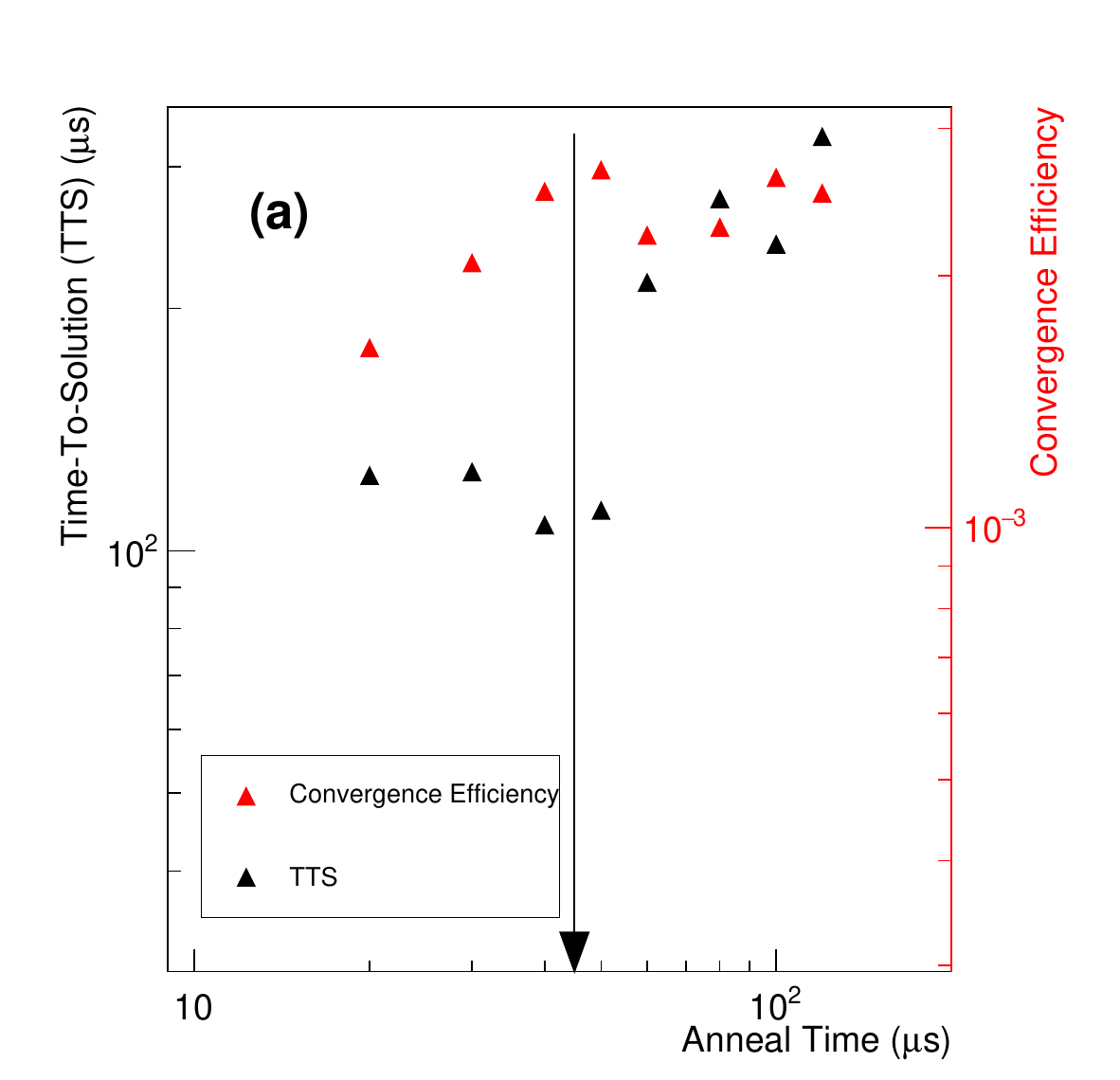} 
\includegraphics[width=\linewidth]{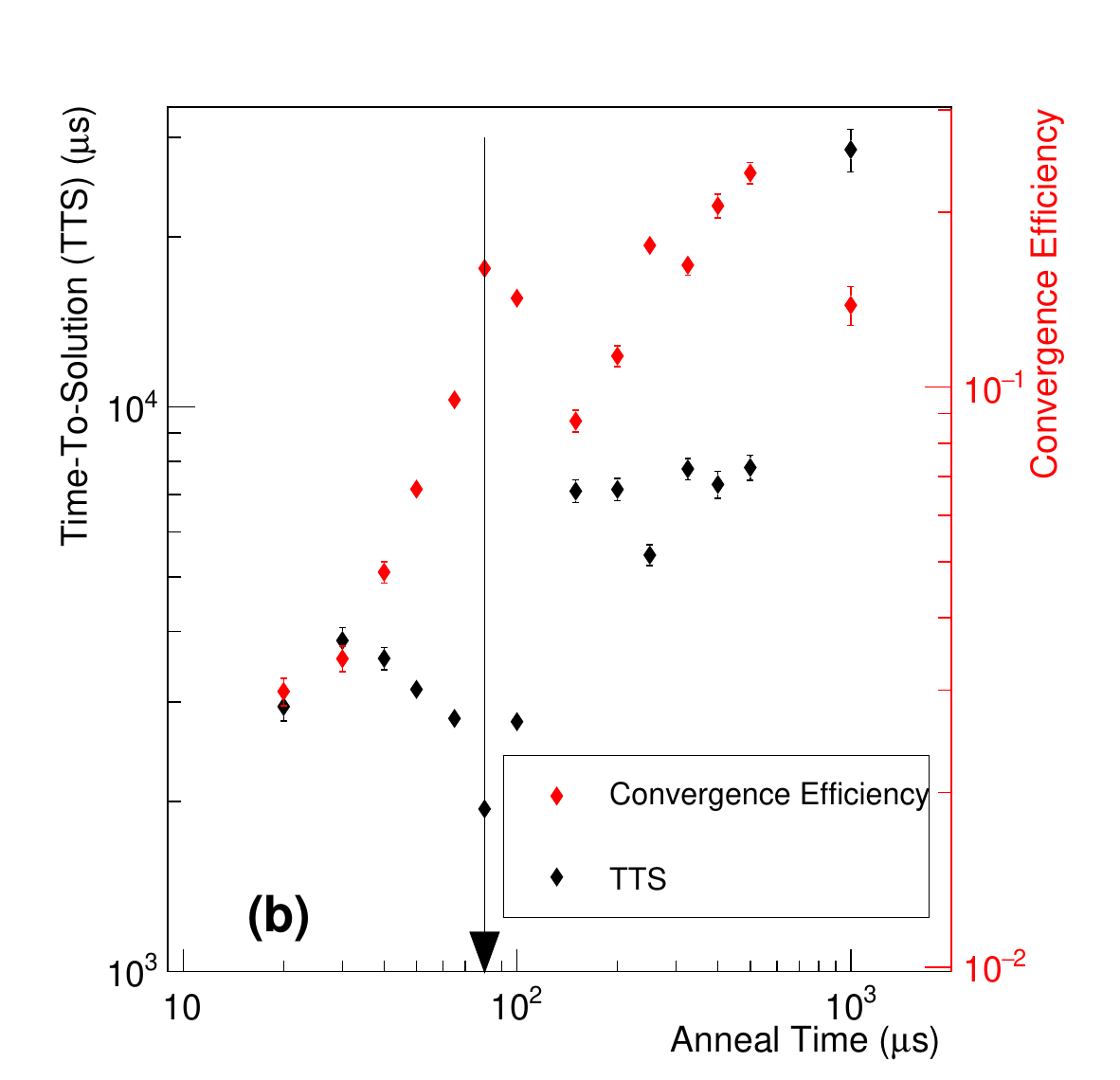} 
\caption{Time to Solution (TTS) and Convergence Efficiency plotted against QPU Annealing Time for two event topologies, (a) 3 vertices 15 tracks, and (b) 5 vertices 15 tracks. Uncertainty bars are statistical, and some are smaller than the markers. Optimal Anneal Time is marked by arrows.}
\label{fig:anneal_time}
\end{figure}

Next, we optimize the annealing time for the QPU as described in Section~\ref{sec:Annealer}. Topologies that result in small gaps in the energy spectrum between the ground state and the next eigenstate need longer annealing times~\cite{ThermallyAssistedQuantumAnnealing}. More complex topologies have denser spectra of eigenstates and therefore smaller spectral gaps. Shorter annealing times offer lower convergence efficiency, and therefore more sampling cycles are needed to find the correct solution. On the other hand, longer annealing times offer higher convergence efficiency, but the total time to obtain a solution increases. The metric we optimize that reflects this trade-off is the Time to Solution (TTS) with a 99\% probability of finding the solution,
\begin{equation}
TTS(t_f) = t_f \log_{1-\epsilon(t_f)}0.01
\label{eq:tts}
\end{equation}
where $t_f$ is the annealing time and $\epsilon$ is the convergence efficiency for that annealing time~\cite{PhysRevX.8.031016}.

We scan the annealing time for two event topologies, 3 vertices 15 tracks and 5 vertices 15 tracks, and identify points of optimal TTS as shown in Fig.~\ref{fig:anneal_time}. Based on this, we set annealing time to 45 $\mu$s for topologies involving less than or equal to 45 logical qubits, and 80 $\mu$s for larger problems. These numbers are important for the following sections on benchmarking.

\section{\label{sec:Benchmark_SA}Benchmarking against Simulated Annealing}

\begin{figure}
\centering
\includegraphics[width=\linewidth]{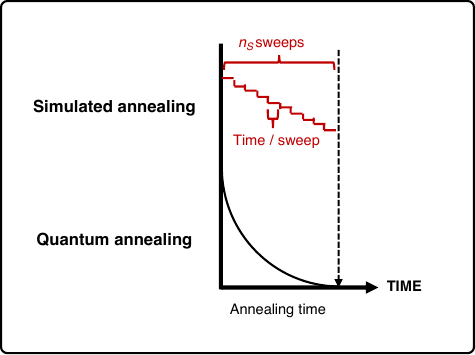} 
\caption{We constrain simulated annealing to use the same time as quantum annealing. For the D-Wave 2000Q quantum annealer, time per sample consists of annealing time, readout time and a re-thermalization delay as described in Section~\ref{sec:OneEvent}. The time / sweep of the simulated annealer is measured as shown in Fig.~\ref{fig:c_TimePerSweep}.}
\label{fig:Benchmarking}
\end{figure}

\begin{figure}
\centering
\includegraphics[width=\linewidth]{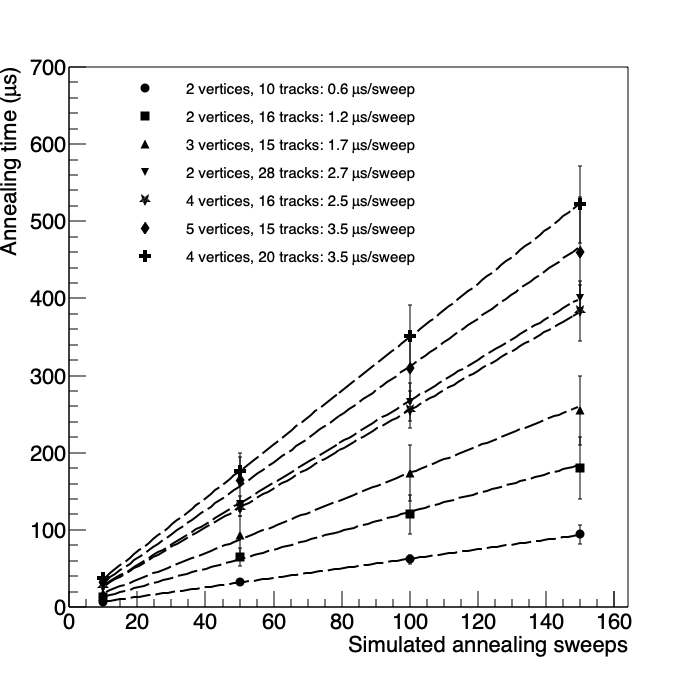}
\caption{Process time of simulated annealing on a 3.1 GHz Intel Core i7-5557U CPU as a function of number of sweeps for various event topologies. The slopes indicate time per sweep for each topology and is used to benchmark against the D-Wave 2000Q\_6 QPU.}
\label{fig:c_TimePerSweep}
\end{figure}

\begin{figure}
\centering
\includegraphics[width=\linewidth]{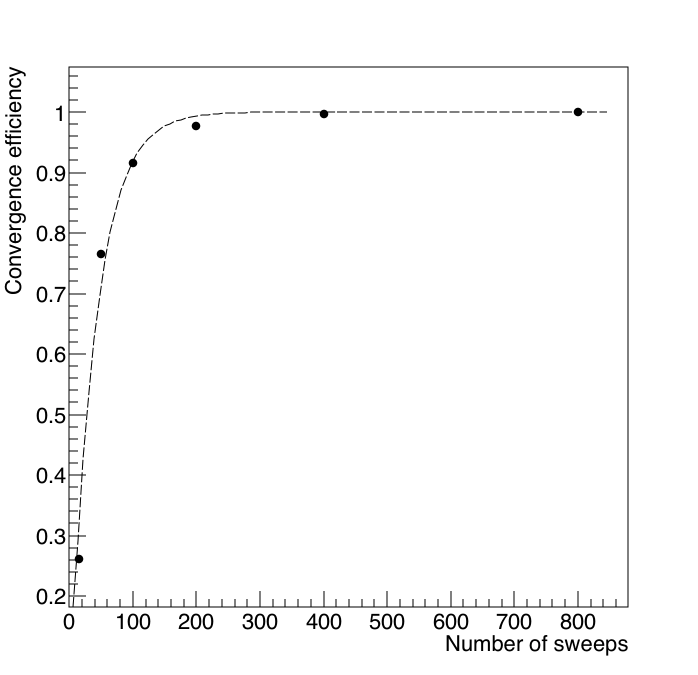}
\caption{Convergence efficiency for simulated annealing to reach the correct answer for clustering an event with 3 primary vertices and 15 tracks as a function of the number of sweeps allowed between initial and final temperatures. Each data point uses 10,000 independent runs of simulated annealing and therefore the statistical uncertainties are negligible. The efficiency asymptotes to one, and is fitted to an exponential charging curve.}
\label{fig:c_SAEfficiencyVsSweeps}
\end{figure}

\begin{figure}
\centering
\includegraphics[width=\linewidth]{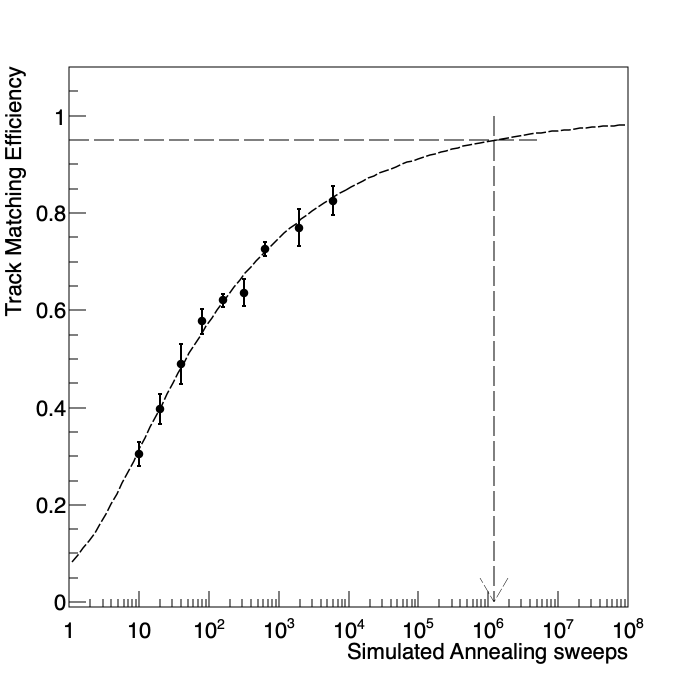}
\caption{Track matching efficiency as a function of sweeps for simulated annealing on one event with 32 vertices and 800 tracks. The number of sweeps necessary to solve for 95\% efficiency is found to be $1.2 \times 10^6$.}
\label{fig:c_LargeProblem}
\end{figure}

We benchmark the D-Wave QPU's performance on this problem against a time-optimized implementation of Simulated Annealing (SA)~\cite{Kirkpatrick671} on one of the cores of a 3.1 GHz Intel Core i7-5557U CPU. The SA algorithm is written for a problem expressed generally in QUBO form for a vector of bits of size $N$ as shown in Eq.~\ref{eq:ProblemHamiltonianQUBO}. The algorithm increments an inverse-temperature parameter $\beta$ linearly from $\beta_i$ to $\beta_f$ in $n_{S}$ number of steps. $\beta_i$ and $\beta_f$ correspond to the initial (hottest) and final (coldest) temperatures relevant to the problem, respectively. We set $\beta_i$ to the order of magnitude of the inverse-temperature at which the most strongly coupled bit has a 50\% probability to flip as shown in Eq.~\ref{eq:beta_i}. This requires us to compute the energy difference due to the flip, $\Delta E_{max}$, if all bits connected to it are 1. By studying all event topologies, we find $\beta_i$ may be reasonably set to 0.1. Similarly, $\beta_f$ is set to the inverse-temperature at which the smallest coupling in the QUBO, $\Delta E_{min}$, has a 1\% chance of a bit flip across it as shown in Eq.~\ref{eq:beta_f}. After studying all topologies, we set $\beta_f$ to 10.
\begin{equation}
    \beta_i = -\log(0.5) / \Delta E_{max}
\label{eq:beta_i}
\end{equation}
\begin{equation}
    \beta_f = -\log(0.01) / \Delta E_{min}
\label{eq:beta_f}
\end{equation}
At each increment in $\beta$, we ``sweep" through the $N$ bits to flip each one at a time. If flipping a bit results in a lower energy, the flip is accepted. If flipping it results in a state with energy $\Delta E$ higher, we accept the flip with probability corresponding to $e^{-\beta\Delta E}$. Else, the flip is reversed.

We compile the algorithm in \texttt{C++} enabling \texttt{-O2} optimization \footnote{The \texttt{-O2} optimization scheme varies between \texttt{C++} compiler flavors but generally avoid execution speed versus memory trade-offs. For Gnu \texttt{C++}, details of \texttt{-O2} optimization may be found here: \href{https://gcc.gnu.org/onlinedocs/gcc/Optimize-Options.html}{https://gcc.gnu.org/onlinedocs/gcc/Optimize-Options.html}}. This was found to be the fastest \texttt{C++} optimization on the CPU. To time-optimize the implementation, we construct a sorted map (\texttt{std::map} in \texttt{C++}) from the QUBO where keys are bit indices $q_i$ and values are the list of bit indices the particular bit is connected to and the couplings between them. This list is encoded in a \mbox{\texttt{std::vector<std::pair \\<unsigned int, double>>}} where the first element of the pair is the other bit $q_j$ it couples to, and the second element is the $b_{ij}$ coupling between them. As a bit flip requires us to only compute the energy difference due to it, this organization streamlines the lookup of the relevant bits and couplings.

Given an infinite number of sweeps, SA will always converge to the ground state by construction. That is to say the system of bits will always jump out of local minima and into the global minimum given sufficient time. To demonstrate this, we run SA on one 3 vertices and 15 tracks event, and plot the efficiency of converging to the correct answer as a function of the number of sweeps in Fig.~\ref{fig:c_SAEfficiencyVsSweeps}. Therefore, to make a fair comparison, we run SA for the number of sweeps that correspond to the annealing time of the D-Wave QPU, repeat both processes a statistically large number of times and compare convergence efficiencies. This requires us to first measure how many SA sweeps ($n_S$) can be accomplished in the annealing time decided upon in Section~\ref{sec:AnnealingTimeOptimization}. For each of our chosen seven topologies, we graph the process time on the CPU against $n_S$ equal to 10, 50, 100, and 150 as shown in Fig.~\ref{fig:c_TimePerSweep}. Thus, we eliminate overhead time and extract the time per sweep from the slope. We note that the time per sweep is linear in problem size ($n_V n_T$) as expected from the memory organization. In Table~\ref{tab:Table_SA}, we record the SA's time per sweep for each topology and infer how many sweeps would fit within the optimal Anneal Time rounded to the nearest integer. Hereafter, we set $n_S$ for SA equal to these numbers of sweeps for each event topology for performance comparison against the quantum annealer.

Can SA on our QUBO formulation be directly used to solve an event with LHC-like numbers of primary vertices? We throw one event with 32 vertices and 800 tracks to find out how long SA takes to solve it on one core of our Intel CPU. By performing a study similar to Fig.~\ref{fig:c_SAEfficiencyVsSweeps}, we find a single sweep for this topology takes 52 ms. Since this is a large event, we define ``Track Matching Efficiency" as the ratio of the number of tracks associated to their correct vertices to the total number of tracks. We then plot the Track Matching Efficiency against number of sweeps in Fig.~\ref{fig:c_LargeProblem}. By fitting an asymptotically rising function, we infer that we need $1.2 \times 10^6$ sweeps to solve for 95\% efficiency. This implies roughly 18 hours. Therefore, it may conceivably be solved within reasonable time if distributed over multiple cores.

\section{\label{sec:Benchmark_DA}Benchmarking against Deterministic Annealing}

Benchmarking against Deterministic Annealing (DA)~\cite{Rose1998} is non-trivial because it is not a sampling algorithm like Simulated or Quantum Annealing. Nevertheless, we use the same principle of restricting the time taken by DA on the same 3.1 GHz Intel Core i7-5557U CPU to the same time taken by a sample of the QPU, and compare convergence efficiencies. The DA algorithm begins by placing all tracks in the same cluster and setting the initial temperature to the critical temperature of that cluster. Next, we begin to lower the temperature iteratively in $n_S$ steps to a preset minimum temperature. At each iteration, we compute the critical temperature of all clusters (which is initially one). If the temperature of the system is found to be lower than the critical temperature of a cluster, that cluster is split into two around its centroid. The probabilities associating each track to a cluster is updated at this stage. At the end of the iterations, the system is frozen to a ``stopping temperature" and the association probabilities are updated.

We set $n_S$ equal to the number of sweeps that correspond to the annealing time of the D-Wave QPU, recorded in Table~\ref{tab:Table_DA}, and compare the efficiency of converging to the correct solution against the quantum annealer.

\begin{figure}
\centering
\includegraphics[width=\linewidth]{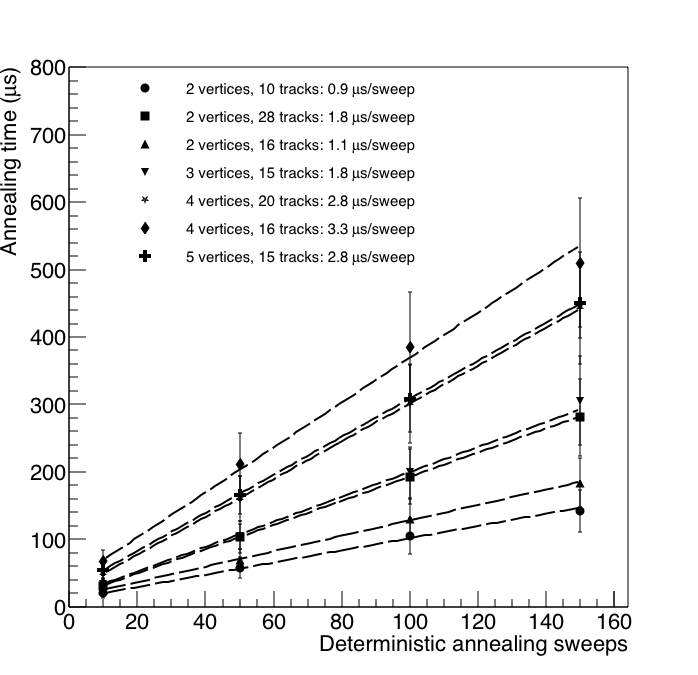}
\caption{Process time of deterministic annealing on a 3.1 GHz Intel Core i7-5557U CPU as a function of number of sweeps for various event topologies. The slopes indicate time per sweep for each topology and is used to benchmark against the D-Wave 2000Q\_6 QPU.}
\label{fig:c_TimePerSweep_DA}
\end{figure}

\section{\label{sec:Results}Results}

First, we apply the formulation, optimizations and benchmarking techniques on one simulated event in Section~\ref{sec:OneEvent}. Second, we apply them on an ensemble of events in Section~\ref{sec:Ensemble}. Finally, we apply our methods on mutiple event topologies to demonstrate how our methods scale with complexity in Section~\ref{sec:Scaling}.

\subsection{\label{sec:OneEvent}Primary vertexing one event}

\begin{figure}
\centering
\includegraphics[width=\linewidth]{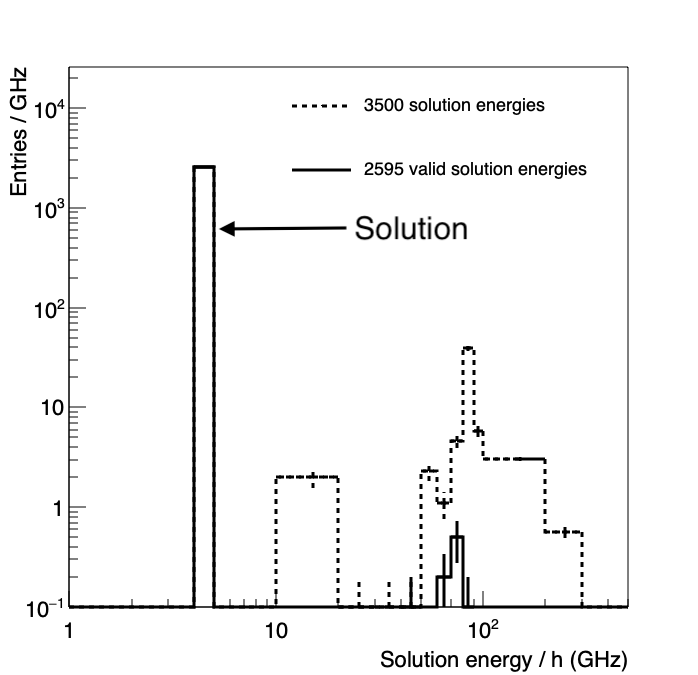}
\caption{Energy spectrum of solutions for one event with 3 primary vertices and 15 tracks explored by the QPU with 10,000 samples. Energies corresponding to valid solutions, where the $p_{ik}$ add up to 1 for every track, are plotted with solid lines while all solutions are plotted with dashed lines. Error bars correspond to statistical uncertainties. For clarity, the histogram is binned by 1 GHz below 10 GHz, by 10 GHz for 10 -- 100 GHz, and by 100 GHz above 100 GHz. Events in 10 (100) GHz bins are normalized by 10 (100).}
\label{fig:c_energy_3V15T_PAPER}
\end{figure}

\begin{figure}
\centering
\includegraphics[width=\linewidth]{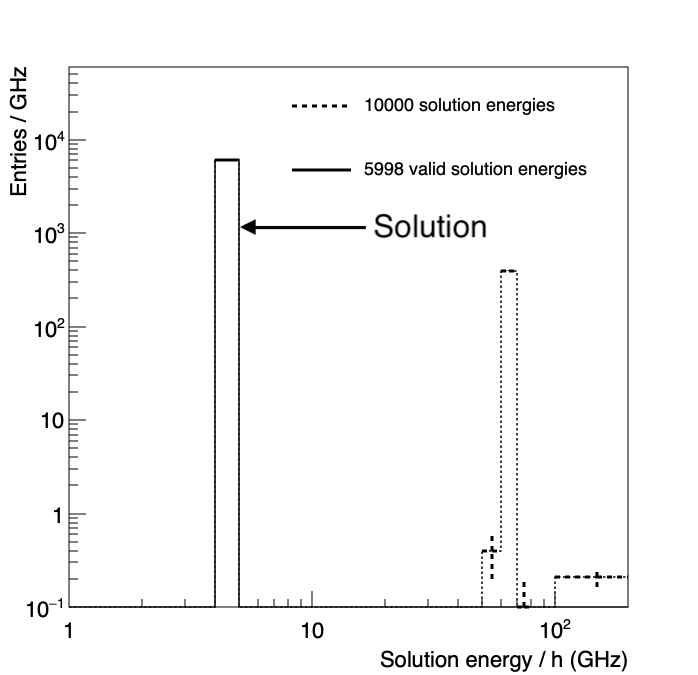}
\caption{Energy spectrum of solutions for one event with 3 primary vertices and 15 tracks explored by SA constrained to 15 sweeps repeated for 10,000 samples. Energies corresponding to valid solutions, where the $p_{ik}$ add up to 1 for every track, are plotted with solid lines while all solutions are plotted with dashed lines. Error bars correspond to statistical uncertainties. The histogram is binned identically to Fig.~\ref{fig:c_energy_3V15T_PAPER}.}
\label{fig:c_energy_SA_3V15T_PAPER}
\end{figure}

To illustrate the algorithm, we generate an event with 3 vertices where 5 tracks emanate from each vertex. This requires 45 logical qubits to encode. The linear and quadratic coefficients associated with the logical qubits are obtained from Eq.~\ref{eq:ObjectiveFunction} in QUBO form.

We sample the QPU 3,500 times to evaluate the efficiency of finding the correct solution. The energy spectrum of the solutions, of which 2,595 are valid (where $p_{ik}$ add up to 1 for every track) is shown in Fig.~\ref{fig:c_energy_3V15T_PAPER}. The energy scale is set by $B(1)$ as defined in Eq.~\ref{eq:Annealing}, which is $12 h$GHz in the QPU used for this study. Of the valid solutions, 2,586 have landed on the lowest energy solution, marked in the figure. On investigating the qubit states, we find that the lowest energy solution corresponds to the correct clustering of the tracks with their respective vertices. Thus, the efficiency of finding the correct solution is noted as 74\%. Higher energy valid solutions correspond to tracks being mis-associated with vertices.

For comparison, the energy spectrum of solutions resulting from SA run 10,000 times with $n_S =$ 27 for each run as discussed in Section~\ref{sec:Benchmark_SA} is shown in Fig.~\ref{fig:c_energy_SA_3V15T_PAPER}. The efficiency of finding valid solutions and the correct solution are noted to both be 60\%.

\subsection{\label{sec:Ensemble}Performance on an ensemble of events}

\begin{figure}
\centering
\includegraphics[width=\linewidth]{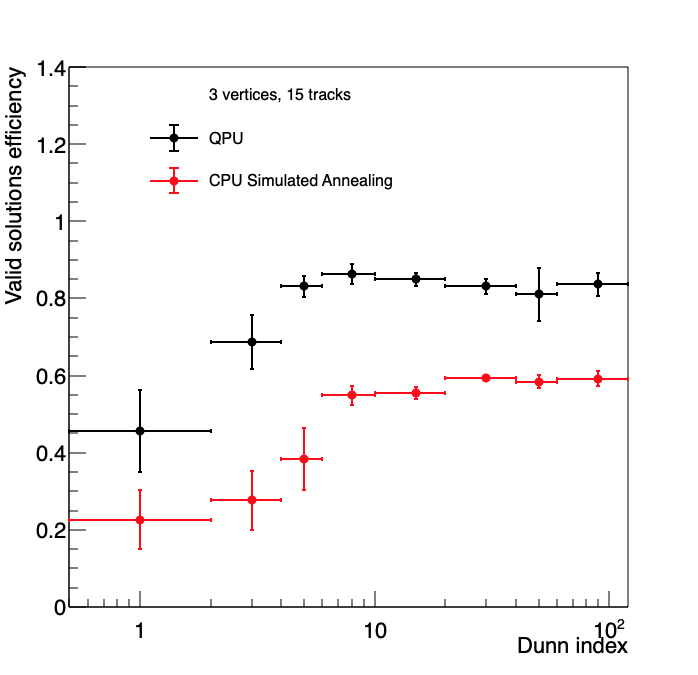}
\caption{Efficiency of finding valid solutions in an ensemble of 100 events with 3 vertices and 15 tracks using the QPU and SA run on the CPU in equal time as a function of event Dunn index.}
\label{fig:c_Dunn_eff_Valid_3V15T_PAPER}
\end{figure}

\begin{figure}
\centering
\includegraphics[width=\linewidth]{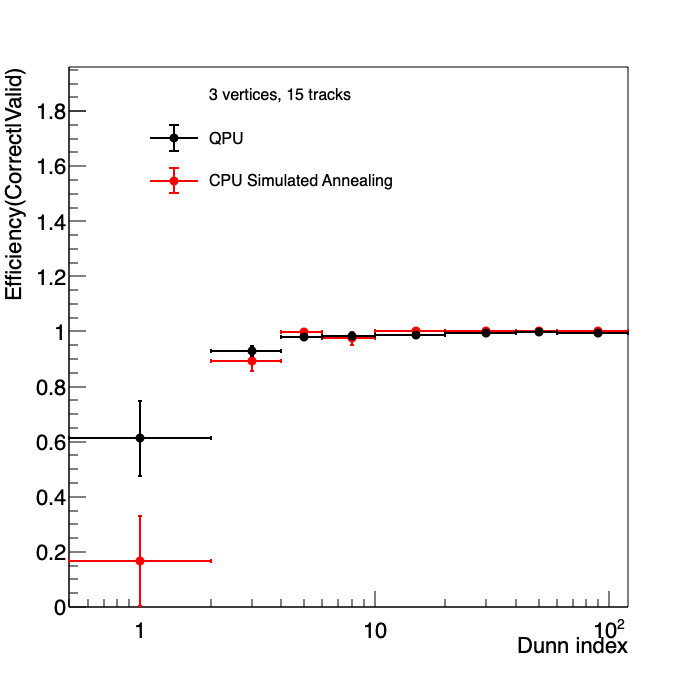}
\caption{Efficiency of finding correct solutions given valid solutions in an ensemble of 100 events with 3 vertices and 15 tracks using the QPU and SA run on the CPU as a function of event Dunn index. The SA is constrained to the anneal time of the QPU.}
\label{fig:c_Dunn_eff_CorrectIfValid_3V15T_PAPER}
\end{figure}

\begin{figure}
\centering
\includegraphics[width=\linewidth]{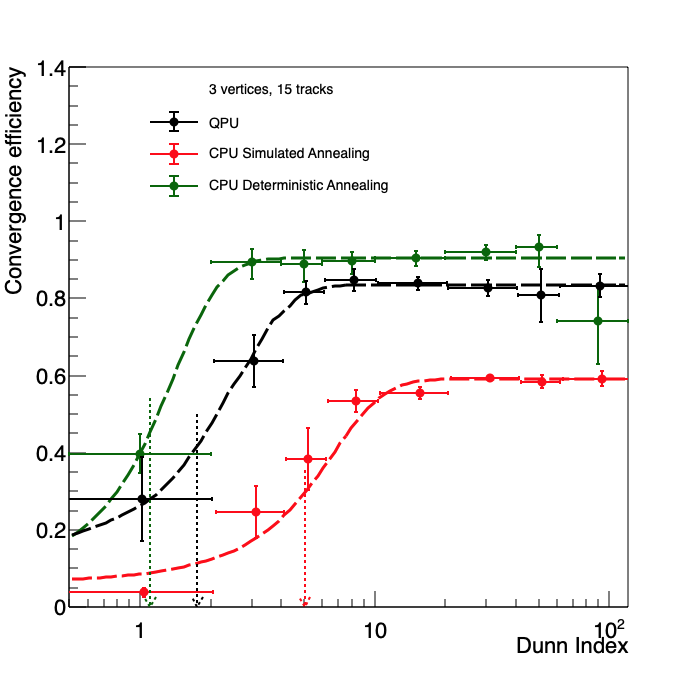}
\caption{Efficiency of finding correct solutions in an ensemble of 100 events with 3 vertices and 15 tracks using the QPU and SA run on the CPU as a function of event Dunn index. The SA is constrained to the anneal time of the QPU.}
\label{fig:c_Dunn_eff_3V15T_PAPER}
\end{figure}

Having studied the performance of the quantum annealer and SA on one particular event, we consider an ensemble of 100 such events with 3 primary vertices and 15 tracks thrown from measured CMS distributions. Events with vertices spaced closely together compared to the spread of their tracks are difficult for the QPU and the SA to solve correctly and result in lower convergence efficiencies than events where vertices are widely separated. We observe this by quantifying the ``clumpiness" of an event by the Dunn index~\cite{Dunn} as discussed below. This also allows us to compare QPU, SA and DA performances in a summary manner.

Since we generate the events, we can characterize the separation of vertices with respect to the intra-vertex spread of tracks with the Dunn index
\begin{equation}
\mathrm{Dunn} = \frac{\mathrm{min_{\emph{k,m}}}\left(d(z_k^V, z_m^V)\right)}{\mathrm{max_{\emph{i,j}}}\left(d(z_i^T, z_j^T)\right)}.
\label{eq:Dunn}
\end{equation}
The numerator of the Dunn index measures the minimum distance between all pairs of primary vertex positions $(z_k^V, z_m^V)$. The denominator measures the maximum distance between all pairs of tracks from one vertex $(z_i^T, z_j^T)$, scanning over all vertices. Thus, a higher Dunn index corresponds to higher clumpiness. Events with higher Dunn indices should be easier for the quantum annealer to reconstruct.

In order to study the dependence of convergence efficiency with Dunn index, we factorize the convergence efficiency into the efficiency of finding valid solutions and the efficiency of finding the correct solutions given valid solutions. To reiterate, a valid solution is one where $p_{ik}$ add up to exactly 1 for every track, i.e. every track must belong to exactly one vertex. In Fig.~\ref{fig:c_Dunn_eff_Valid_3V15T_PAPER}, we show the dependence of valid solutions efficiency on the Dunn index for the QPU and the SA being run on the CPU. As discussed in Section~\ref{sec:Benchmark_SA}, the SA is restricted to the QA's anneal time. The x-axis is placed on a logarithmic scale to highlight the earlier rise of the QPU relative to the CPU, in addition to its higher asymptotic efficiency. The y-uncertainty in each bin corresponds to the standard error on the mean of event efficiencies within it. The statistical uncertainty of each event efficiency is negligible because we take 3,500 and 10,000 samples per event for the QPU and SA, respectively. The x-uncertainties correspond to chosen bin widths.

We plot the dependence of efficiency for finding correct solutions given valid solutions against the Dunn index in Fig.~\ref{fig:c_Dunn_eff_CorrectIfValid_3V15T_PAPER}. This is found to be more or less fully efficient except for low Dunn indices. These two plots now allow us to understand Fig.~\ref{fig:c_Dunn_eff_3V15T_PAPER} which is their product -- the overall efficiency of finding correct solutions as a function of Dunn index. The uncertainties for the QPU and the CPU Simulated Annealing graphs are calculated as described earlier. In this plot, we also show the efficiency of Deterministic Annealing on the same events with constraints discussed in Section~\ref{sec:Benchmark_DA}. Since DA is not a sampling procedure and returns either a success or failure for an event, we calculate the efficiency in each Dunn index bin as the ratio of number of successes to events in that bin. The uncertainty is marked by the 95\% Clopper-Pearson interval~\cite{ClopperPearson}.

To quantify the dependence on Dunn index, we fit the data points with the generic logistic function in Eq.~\ref{eq:Logistic}, where $L$ is the asymptotic efficiency, $x_0$ is the inflection point and $k$ is the slope at the inflection point. The positions of the inflection points in Fig.\ref{fig:c_Dunn_eff_3V15T_PAPER} are indicated with dashed vertical arrows to guide the eye. 
\begin{equation}
y = \frac{L}{1+e^{-k(x-x_0)}}
\label{eq:Logistic}
\end{equation}
The QPU, CPU SA and CPU DA achieve asymptotic efficiencies of 0.84 $\pm$ 0.01, 0.59 $\pm$ 0.00 and 0.91 $\pm$ 0.05, respectively. The inflection point in Dunn index for QPU, CPU SA and CPU DA are 1.8 $\pm$ 1.0, 5 $\pm$ 1 and 1.4 $\pm$ 0.9, respectively. Thus, for this event topology, one can infer a quantum advantage of the QPU over CPU Simulated Annealing, both in terms of asymptotic efficiency and inflection point.

\subsection{\label{sec:Scaling}Scaling with event complexity}

\begin{figure*}
\centering
\includegraphics[width=0.4\linewidth]{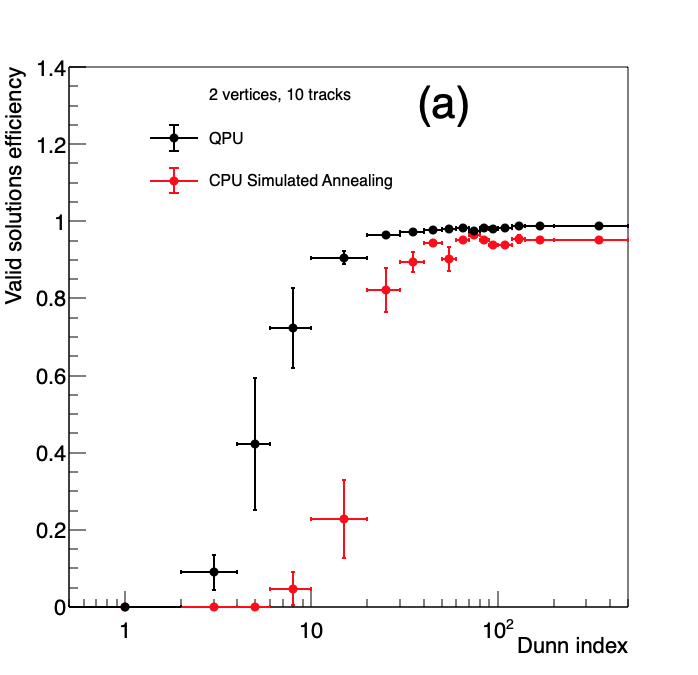}
\includegraphics[width=0.4\linewidth]{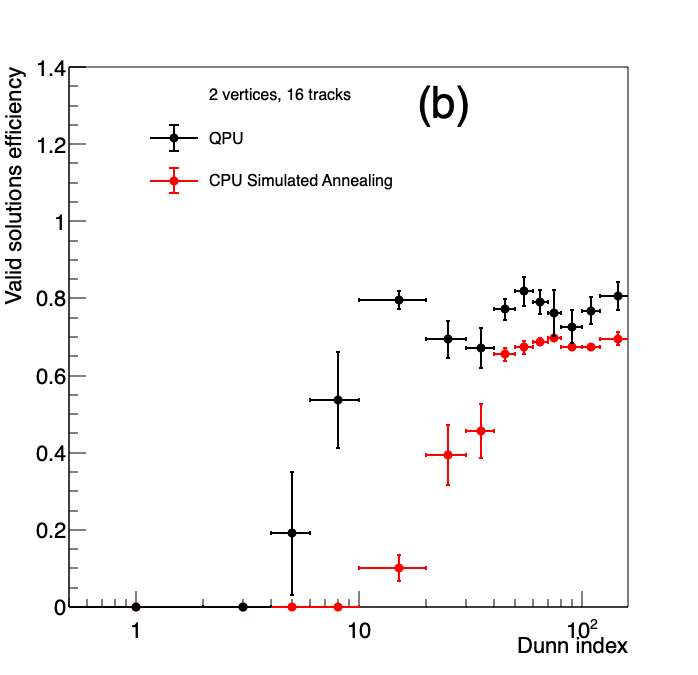}
\includegraphics[width=0.4\linewidth]{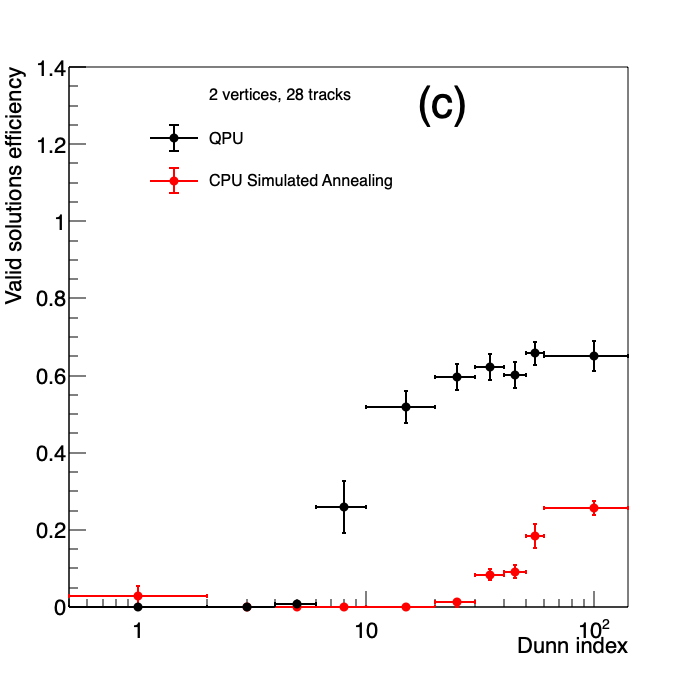}
\includegraphics[width=0.4\linewidth]{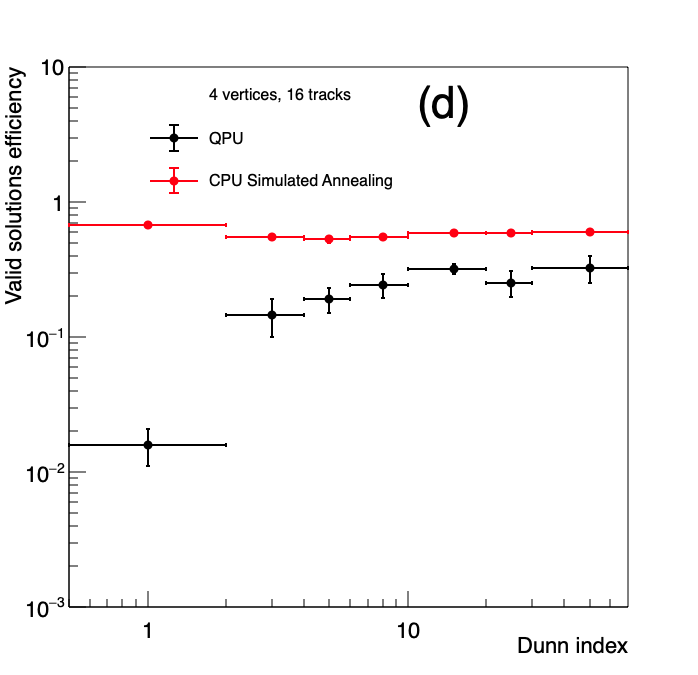}
\includegraphics[width=0.4\linewidth]{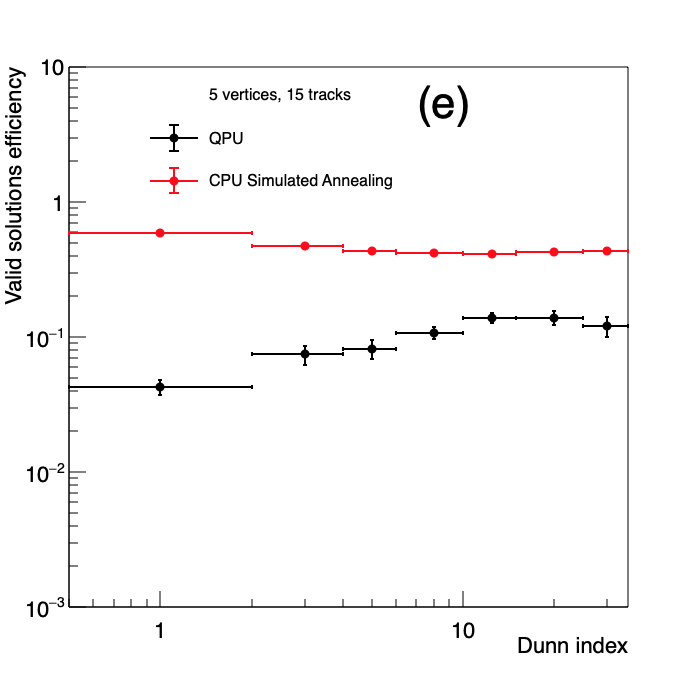}
\includegraphics[width=0.4\linewidth]{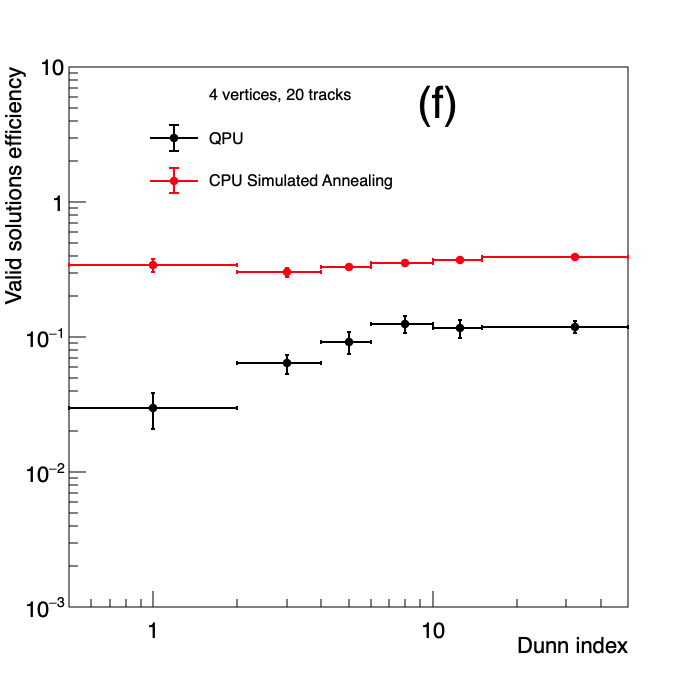}
\caption{Efficiency of finding valid solutions as a function of event Dunn index for various event topologies. Results from the QPU (black) are overlaid with results from Simulated Annealing on the CPU (red) executed in comparable time. Uncertainties are described in Section~\ref{sec:Ensemble}.}
\label{fig:ValidEfficiency_Dunn}
\end{figure*}

\begin{figure*}
\centering
\includegraphics[width=0.4\linewidth]{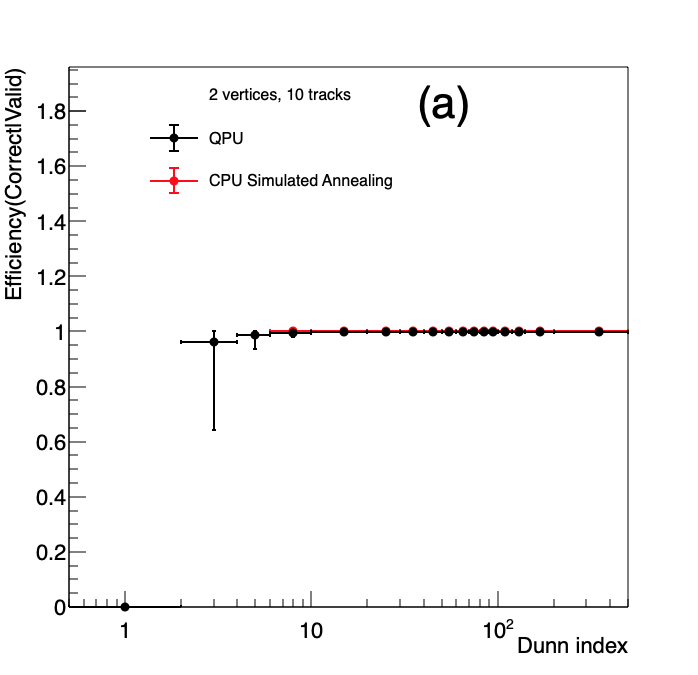}
\includegraphics[width=0.4\linewidth]{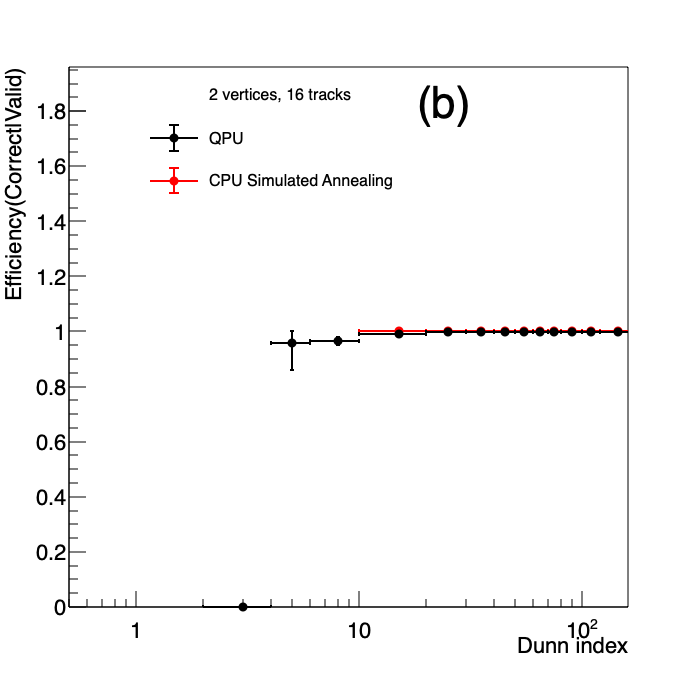}
\includegraphics[width=0.4\linewidth]{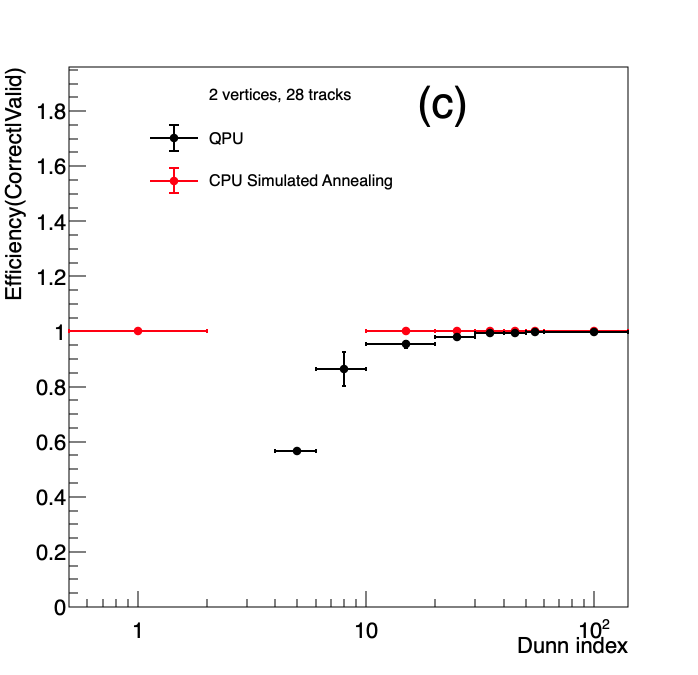}
\includegraphics[width=0.4\linewidth]{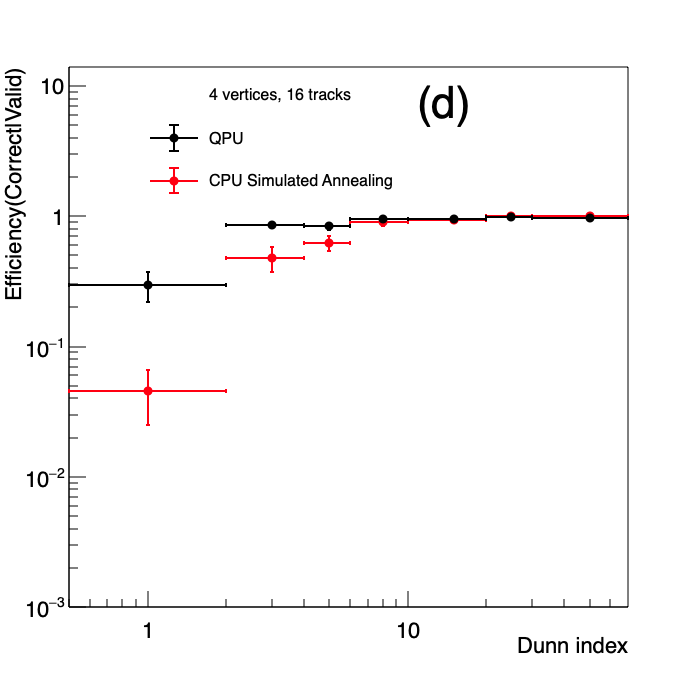}
\includegraphics[width=0.4\linewidth]{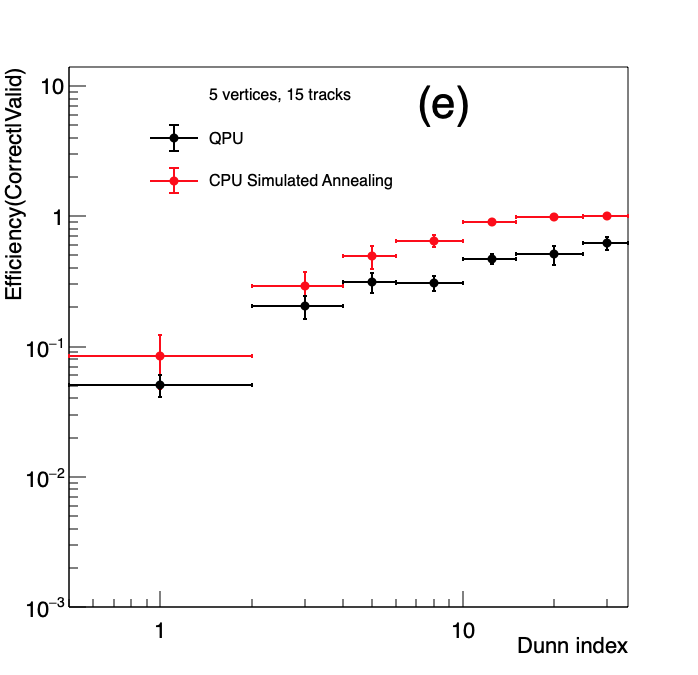}
\includegraphics[width=0.4\linewidth]{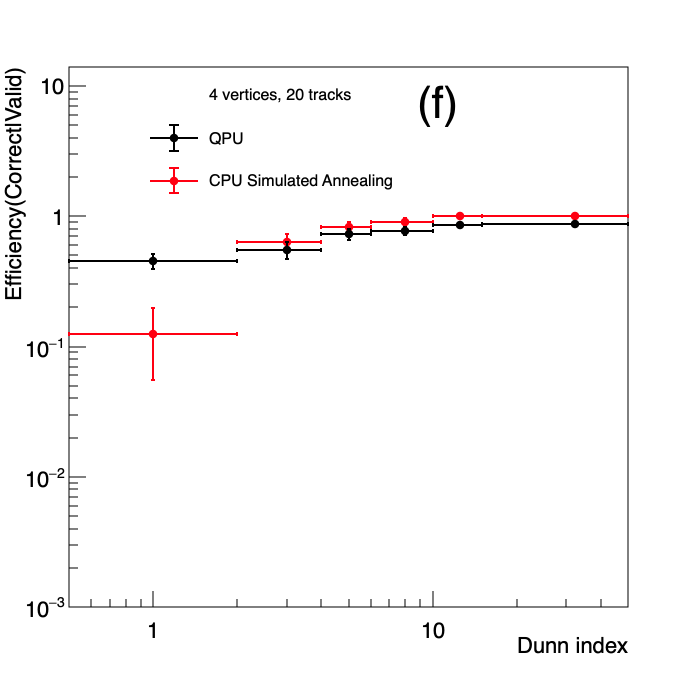}
\caption{Efficiency of finding correct solutions given valid solutions as a function of event Dunn index for various event topologies. Results from the QPU (black) are overlaid with results from Simulated Annealing on the CPU (red) executed in comparable time. Uncertainties are described in Section~\ref{sec:Ensemble}.}
\label{fig:CorrectIfValidEfficiency_Dunn}
\end{figure*}

\begin{figure*}
\centering
\includegraphics[width=0.4\linewidth]{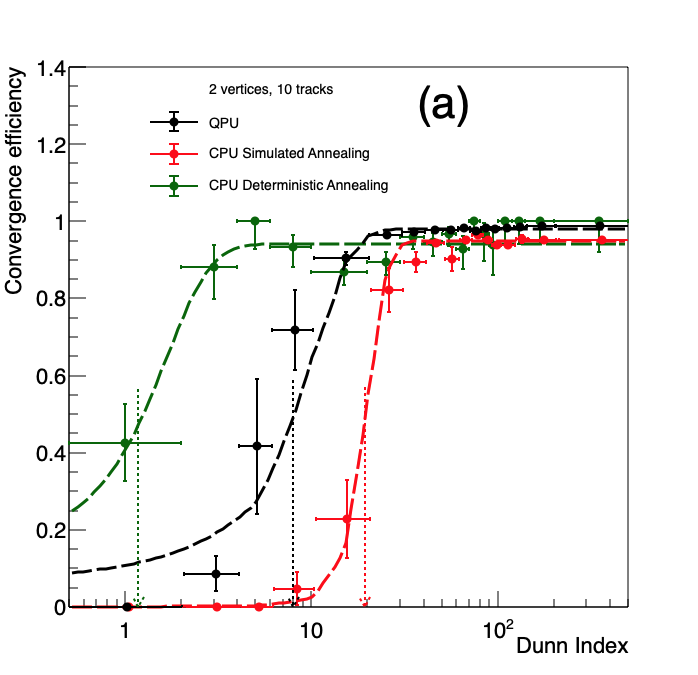}
\includegraphics[width=0.4\linewidth]{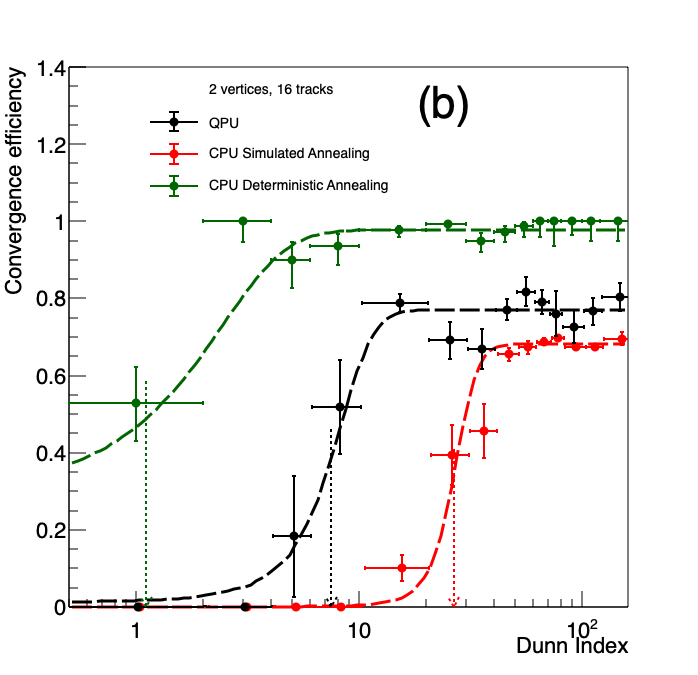}
\includegraphics[width=0.4\linewidth]{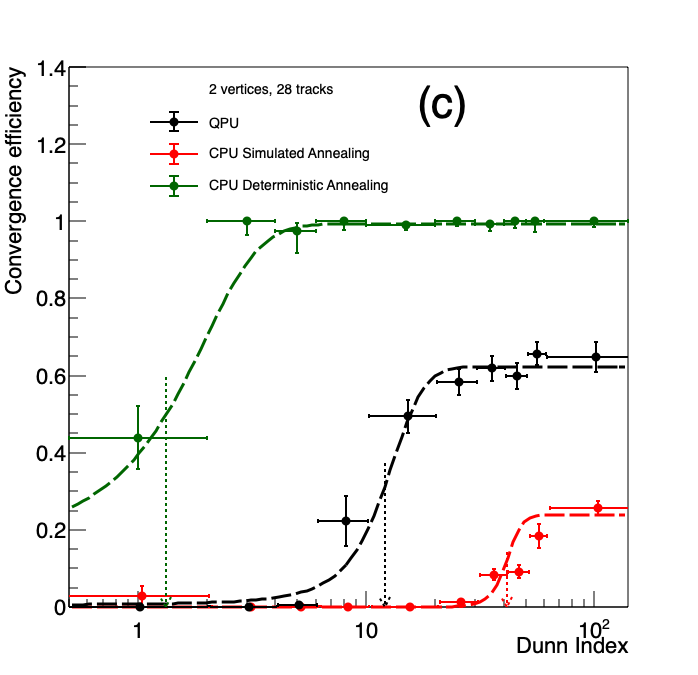}
\includegraphics[width=0.4\linewidth]{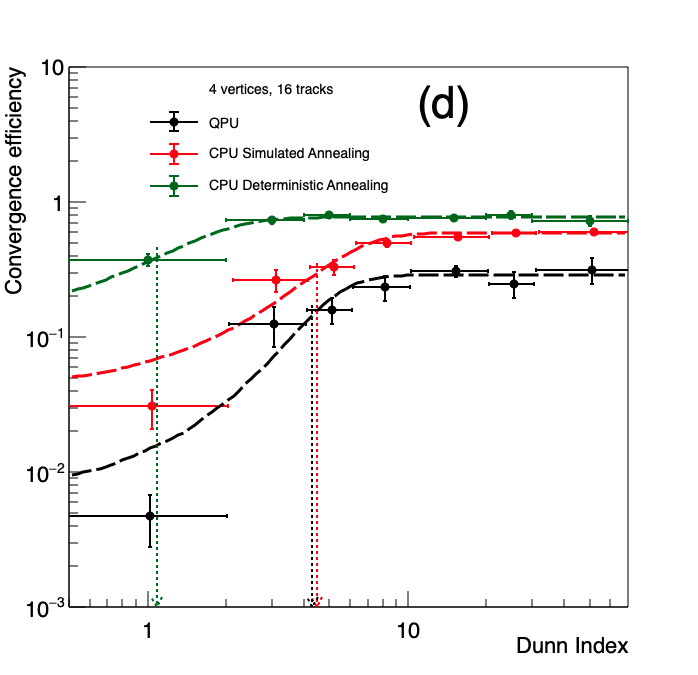}
\includegraphics[width=0.4\linewidth]{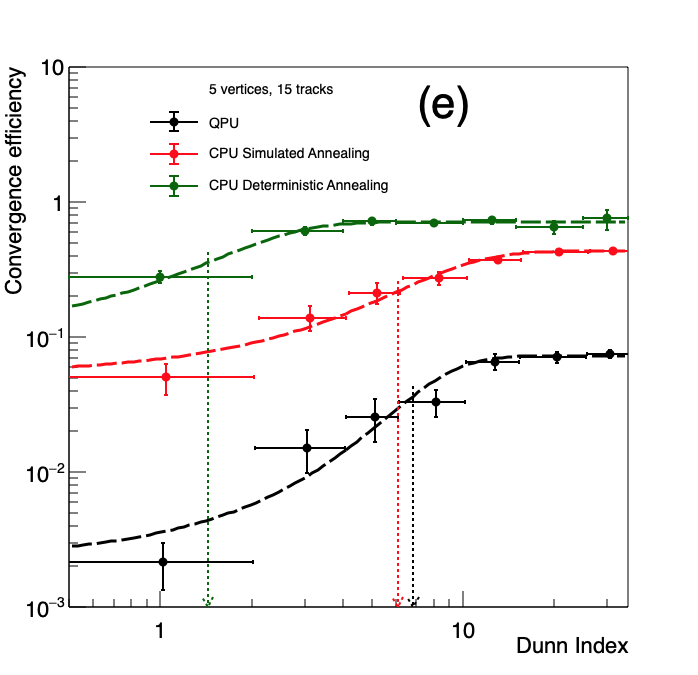}
\includegraphics[width=0.4\linewidth]{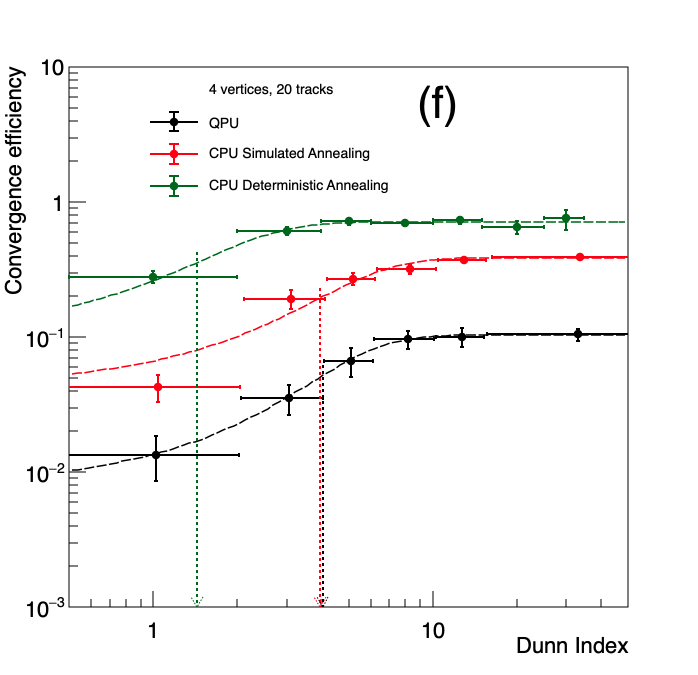}
\caption{Efficiency of finding correct solutions as a function of event Dunn index for various event topologies. Results from the QPU (black) are overlaid with results from Simulated Anneailng (red) and Deterministic Annealing (green) executed in comparable time. Uncertainties and fits are described in Section~\ref{sec:Ensemble}.}
\label{fig:Efficiency_Dunn}
\end{figure*}

\begin{table*}
\centering
\begin{tabular}{c c c c}
\hline
\hline
          & \multicolumn{3}{c}{Asymptotic Efficiency} \\
Topology & QPU & CPU SA & CPU DA \\
\hline
2 vertices 10 tracks & 0.98 $\pm$ 0.00 & 0.95 $\pm$ 0.00 & 0.94 $\pm$ 0.01 \\
2 vertices 16 tracks & 0.77 $\pm$ 0.01 & 0.68 $\pm$ 0.00 & 0.98 $\pm$ 0.01 \\
3 vertices 15 tracks & 0.84 $\pm$ 0.01 & 0.59 $\pm$ 0.00 & 0.91 $\pm$ 0.01 \\
2 vertices 28 tracks & 0.62 $\pm$ 0.02 & 0.24 $\pm$ 0.02 & 0.99 $\pm$ 0.00 \\
4 vertices 16 tracks & 0.29 $\pm$ 0.02 & 0.59 $\pm$ 0.01 & 0.77 $\pm$ 0.02 \\
5 vertices 15 tracks & 0.073 $\pm$ 0.004 & 0.434 $\pm$ 0.006 & 0.71 $\pm$ 0.02 \\
4 vertices 20 tracks & 0.104 $\pm$ 0.009 & 0.387 $\pm$ 0.006 & 0.71 $\pm$ 0.02 \\
\hline
\hline
\end{tabular}
\caption{Asymptotic efficiencies of the QPU, the CPU Simulated Annealing, and the CPU Deterministic Annealing estimated from the fits in Fig.~\ref{fig:Efficiency_Dunn} and Fig.~\ref{fig:c_Dunn_eff_3V15T_PAPER}.}
\label{tab:Table_Convergence}
\end{table*}

\begin{table*}
\centering
\begin{tabular}{c c c c}
\hline
\hline
          & \multicolumn{3}{c}{Inflection Point Dunn Index} \\
Topology & QPU & CPU SA & CPU DA \\
\hline
2 vertices 10 tracks & 8.0 $\pm$ 0.3 & 19   $\pm$ 8  & 1.2 $\pm$ 0.9 \\
2 vertices 16 tracks & 7.5 $\pm$ 1.6 & 27   $\pm$ 2   & 1.1 $\pm$ 1.0 \\
3 vertices 15 tracks & 1.8 $\pm$ 0.8 & 5    $\pm$ 1   & 1.1 $\pm$ 0.9 \\
2 vertices 28 tracks & 12  $\pm$ 2   & 41  $\pm$ 1    & 1.3 $\pm$ 1.3 \\
4 vertices 16 tracks & 4.3   $\pm$ 0.9   & 4.5 $\pm$ 0.7  & 1.1 $\pm$ 0.8 \\
5 vertices 15 tracks & 6.8 $\pm$ 1.2     & 6.1 $\pm$ 0.8      & 1.4 $\pm$ 0.8 \\
4 vertices 20 tracks & 4 $\pm$ 1     & 4 $\pm$ 1      & 1.4 $\pm$ 0.8 \\
\hline
\hline
\end{tabular}
\caption{Inflection points in Dunn index of the QPU, the CPU Simulated Annealing, and the CPU Deterministic Annealing estimated from the fits in Fig.~\ref{fig:Efficiency_Dunn} and Fig.~\ref{fig:c_Dunn_eff_3V15T_PAPER}.}
\label{tab:Table_Inflection}
\end{table*}

\begin{table*}
\centering
\begin{tabular}{c c c c}
\hline
\hline
          & \multicolumn{3}{c}{Slope (Efficiency / Dunn index)} \\
Topology & QPU & CPU SA & CPU DA \\
\hline
2 vertices 10 tracks & 0.33 $\pm$ 0.03 & 0.40 $\pm$ 0.01 & 1.6 $\pm$ 1.1 \\
2 vertices 16 tracks & 0.60 $\pm$ 0.04 & 0.30 $\pm$ 0.01 & 0.8 $\pm$ 0.4 \\
3 vertices 15 tracks & 1.10 $\pm$ 0.47  & 0.45 $\pm$ 0.08 & 2.3 $\pm$ 2.7 \\
2 vertices 28 tracks & 0.40 $\pm$ 0.01 & 0.30 $\pm$ 0.01 & 1.3 $\pm$ 0.7 \\
4 vertices 16 tracks & 0.9 $\pm$ 0.5   & 0.6  $\pm$ 0.2  & 1.6 $\pm$ 1.1 \\
5 vertices 15 tracks & 0.5 $\pm$ 0.1   & 0.33 $\pm$ 0.07 & 1.3 $\pm$ 0.8 \\
4 vertices 20 tracks & 0.6 $\pm$ 0.3   & 0.5 $\pm$ 0.1   & 1.3 $\pm$ 0.8 \\
\hline
\hline
\end{tabular}
\caption{Slope at the inflection points in Dunn index of the QPU, the CPU Simulated Annealing, and the CPU Deterministic Annealing estimated from the fits in Fig.~\ref{fig:Efficiency_Dunn} and Fig.~\ref{fig:c_Dunn_eff_3V15T_PAPER}.}
\label{tab:Table_Slope}
\end{table*}

To characterize how convergence efficiency scales with event complexity, we repeat our investigation for the six other event topologies mentioned in Section~\ref{sec:Embedding} and presented in Table~\ref{tab:Table_SA}. Each of the topologies are deterministically embedded on the Chimera architecture, as discussed in Section~\ref{sec:Embedding}, which gives us the average chain lengths listed in the table. Further, the chain strengths were optimized for each of the topologies as discussed in Section~\ref{sec:ChainStrengthOptimization}. The dependence of valid solutions efficiency on Dunn index for each of these topologies, for both QPU and CPU SA, is shown in Fig.~\ref{fig:ValidEfficiency_Dunn}. The ratio of correct solutions to valid solutions as a function of Dunn index is shown in Fig.~\ref{fig:CorrectIfValidEfficiency_Dunn}. 

The aforementioned sets of plots may be multiplied, point by point, to obtain the efficiency for obtaining correct solutions as a function of Dunn index as shown in the set of plots in Fig.~\ref{fig:Efficiency_Dunn}. Results for CPU DA are also overlaid on these plots. A logarithmic scale is used for the y-axis where efficiencies are dramatically different. Fits to the logistic function are overlaid with inflection points marked. Results for the asymptotic efficiency, inflection point, and slope are recorded in Tables~\ref{tab:Table_Convergence}, \ref{tab:Table_Inflection} and \ref{tab:Table_Slope}, respectively.

\begin{figure}
\centering
\includegraphics[width=\linewidth]{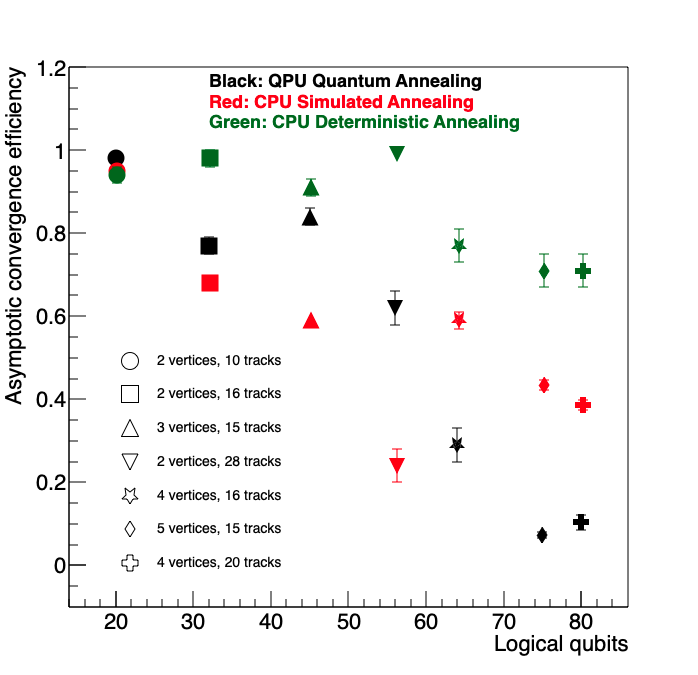}
\caption{The dependence of QPU (black), SA with CPU (red) and DA with CPU (green) asymptotic convergence efficiency on track clustering problem complexity measured by the number of logical qubits needed. The processor time spent on the CPU per sample is equal to that spent by the QPU per sample, for all topologies. The values are from Table~\ref{tab:Table_Convergence}.}
\label{fig:c_MaxConvEff_qubits_error_PAPER}
\end{figure}

We summarize the performances of the QPU against the CPU SA and DA in Fig.~\ref{fig:c_MaxConvEff_qubits_error_PAPER} by plotting the asymptotic efficiency against the number of logical qubits needed to encode the studied event topologies. We note that while performances are comparable for the 2 vertices 10 tracks topology, quantum annealing holds an advantage over simulated annealing up to the topology of 2 vertices 28 tracks (56 logical qubits). Thereafter, the QPU's performance drops dramatically for the 4 vertices 16 tracks topology. Explanations for this drop-off are necessarily speculative at this stage. One explanation could be that the increased density of eigenstates near the ground-state cause more frequent anti-crossings to higher energy states. Another could be that the long chains simply freeze out faster into local minima. Addressing these will need detailed investigations and can motivate further research as outlined in Section~\ref{sec:Conclusions}.

\section{\label{sec:Conclusions}Conclusions and outlook}

With noisy intermediate-scale quantum computers commercially available, we demonstrate that a D-Wave 2000Q\_6 quantum annealer can perform track clustering for primary vertexing at hadron collider experiments. We have shown that a quantum advantage (slight edge) persists over Simulated Annealing run on a commercial CPU for 2 vertices 10 tracks, 2 vertices 16 tracks, 3 vertices 15 tracks, and 2 vertices 28 tracks that would have been relevant at a Tevatron experiment like D0. Along the way, we have demonstrated a deterministic graph-embedding of the clustering formulation on the D-Wave Chimera architecture, a method for optimizing chain strengths, a method for optimizing annealing time, and a paradigm for benchmarking performance against simulated annealing. We also found a linear relationship between chain lengths and optimal chain strengths that can be exploited by future researchers.

To solve problem sizes of 35 to 200 primary vertices relevant for the LHC will likely require using this formulation as an algorithmic primitive. One may use the algorithm to first cluster tracks into 2 or 3 vertices and then sub-cluster them into more vertices recursively till an energy threshold is reached. Optimizing the parameters of this hierarchical algorithm, like the number of clusters to form in each iteration and how to define the stopping point, is beyond the scope of this paper. However, the most useful approach to solving LHC-scale primary vertexing will be to:

\begin{enumerate}
\item Adapt the formulation, embedding and optimization methods outlined in this paper for the D-Wave Advantage processor with the Pegasus architecture, which has over 5,000 qubits and 15 couplers per qubit. The Zephyr architecture with 20 couplers per qubit is also on the horizon.
\item Investigate the cause of efficiency drop-offs for high event complexities. If this is caused by long chains freezing out, it may be remedied by optimizing the annealing schedule per chain~\cite{Lanting2017}. If anti-crossing is the culprit, methods to increase the spectral gap must be developed that are similar in spirit to but more sophisticated than our distortion function $g(x; m)$. Modifying the annealing schedule~\cite{ReverseAnnealing1, ReverseAnnealing2}, and introducing pauses~\cite{Pausing1, Pausing2} are also tools that may be used once the cause is understood.
\item Develop the aforementioned hierarchical algorithm that uses our formulation adapted and optimized for a next-generation processor as a primitive.
\end{enumerate}

We hope that the methods outlined in this paper will serve as a springboard for further research into the formulation and optimization of clustering algorithms relevant for high energy physics on quantum annealers beyond the narrow scope of primary vertexing.

\section*{Acknowledgements}

We acknowledge several useful discussions with Sachin B. Vaidya in the initial stages of this project. We also acknowledge useful discussions with Joel Gottlieb, Mark Johnson, and Alexander Condello at D-Wave Systems Inc., and are grateful for the opportunity to work with multiple versions of the 2000Q quantum annealer. The authors acknowledge support from the Purdue Research Foundation and the Purdue Quantum Science and Engineering Institute Seed Grant.

\bibliography{TrackClusteringDWave_NIM}

\end{document}